\documentclass[amsmath,nobibnotes,aps,superscriptadress,amssymb,pra,aps,showpacs,superscriptaddress,twocolumn]{revtex4-1}
\usepackage{amsmath,amsfonts,amssymb,amsthm,graphics,graphicx,epsfig,bbm}
\usepackage[colorlinks=true,citecolor=blue,linkcolor=blue,urlcolor=blue]{hyperref}
\usepackage[usenames]{color}
\usepackage{adjustbox}

\usepackage{float}
\makeatletter
\let\newfloat\newfloat@ltx
\makeatother

\usepackage{algorithm}
\usepackage{algpseudocode}

\usepackage{algorithmicx}
\usepackage{graphicx}
\usepackage{tikz}
\usepackage{subfigure}
\usepackage{amsmath}
\usepackage{epsfig}
\usepackage{dcolumn}
\usepackage{bm}
\usepackage{color}
\usepackage{times}
\usepackage{epstopdf}
\usepackage{amscd}
\usepackage{amsthm}
\usepackage{amssymb}
\usepackage{bbm}
\usepackage{amssymb}
\usepackage{amstext}
\usepackage{latexsym}
\usepackage{hyperref}
\usepackage{amsfonts}
\usepackage{psfrag}
\usepackage{soul,xcolor}
\usepackage[normalem]{ulem}
\usepackage{dsfont}
\usepackage{txfonts}
\usepackage{physics}
\usepackage{footnote}
\usepackage{multirow}
\usepackage{appendix}
\usepackage{mathtools}
\usepackage{tikz}
\usetikzlibrary{quantikz}%to draw quantum circuits
\usepackage{bbold}
\usepackage{enumitem}
\usepackage{xstring}
\usepackage{cancel}
\usepackage{amscd}
\usepackage{bbm}
\usepackage{units}
\usepackage{cancel}
\usepackage{xstring}
\usepackage{import}
\usepackage{subfiles}

\definecolor{mycolor}{RGB}{106,81,162}
\newtheorem{th1}{Theorem}

\newtheorem{cor1}{Corollary}

\newtheorem{de1}{Definition}

\usepackage{mathtools}

\newcommand{\npartial}[2]{\partial_{#1}^{(#2)}}
\newcommand{\iden}[1]{
    \ifthenelse{\equal{1}{\string #1}}
  {% True case
   \mathbbm{1}
  }
  {% false case
   \mathbbm{1}^{\otimes#1}}
  }
\newcommand{\ketzero}[1]{
    \ifthenelse{\equal{1}{\string #1}}
  {% True case
   \ket{0}
  }
  {% false case
   \ket{0}^{\otimes#1}}
  }
\newcommand{\brazero}[1]{
    \ifthenelse{\equal{1}{\string #1}}
  {% True case
   \bra{0}
  }
  {% false case
   \bra{0}^{\otimes#1}}
  }
\newcommand{\ketone}[1]{
      \ifthenelse{\equal{1}{\string #1}}
    {% True case
     \ket{1}
    }
    {% false case
     \ket{1}^{\otimes#1}}
    }
  \newcommand{\braone}[1]{
      \ifthenelse{\equal{1}{\string #1}}
    {% True case
     \bra{1}
    }
    {% false case
     \bra{1}^{\otimes#1}}
    }
\definecolor{mycolor}{RGB}{106,81,162}

\theoremstyle{plain}
\newtheorem{thm}{Theorem}

\newtheorem{lem}{Lemma}

\theoremstyle{definition}
\newtheorem{defn}{Definition}
\theoremstyle{remark}

\begin{document}

\setstcolor{red}
\newtheorem{Proposition}{Proposition}[section]

\title{Quantum algorithms for approximate function loading}
\date{\today}

\author{Gabriel Marin-Sanchez }
\affiliation{Department of Physical Chemistry, University of the Basque Country UPV/EHU, Apartado 644, 48080 Bilbao, Spain}

\author{Javier Gonzalez-Conde}
\email[Corresponding author: ]{\qquad javier.gonzalezc@ehu.eus}
\affiliation{Department of Physical Chemistry, University of the Basque Country UPV/EHU, Apartado 644, 48080 Bilbao, Spain}
\affiliation{EHU Quantum Center, University of the Basque Country UPV/EHU, Apartado 644, 48080 Bilbao, Spain}

%\affiliation{Quantum Mads, Uribitarte Kalea 6, 48001 Bilbao, Spain}

\author{Mikel Sanz}
\email[Corresponding author: ]{\qquad mikel.sanz@ehu.eus}
\affiliation{Department of Physical Chemistry, University of the Basque Country UPV/EHU, Apartado 644, 48080 Bilbao, Spain}
\affiliation{EHU Quantum Center, University of the Basque Country UPV/EHU, Apartado 644, 48080 Bilbao, Spain}
\affiliation{IKERBASQUE, Basque Foundation for Science, Plaza Euskadi 5, 48009, Bilbao, Spain}
\affiliation{Basque Center for Applied Mathematics (BCAM),  Alameda de Mazarredo, 14, 48009 Bilbao, Spain}
\begin{abstract}

Loading classical data into quantum computers represents an essential stage in many relevant quantum algorithms, especially in the field of quantum machine learning. Therefore, the inefficiency of this loading process means a major bottleneck for the application of these algorithms. Here, we introduce two approximate quantum-state preparation methods  for the NISQ era inspired by the Grover-Rudolph algorithm, which partially solve the problem of loading real functions. Indeed, by allowing for an infidelity $\epsilon$ and under certain smoothness conditions, we prove that the complexity of the implementation of the Grover-Rudolph algorithm without ancillary qubits, first introduced by  Möttönen \textit{et al}, results into $\mathcal{O}(2^{k_0(\epsilon)})$, with $n$ the number of qubits and $k_0(\epsilon)$ asymptotically independent of $n$. This leads to a dramatic reduction in the number of required two-qubit gates. Aroused by this result, we also propose a variational algorithm capable of loading functions beyond the aforementioned smoothness conditions. Our variational Ansatz is explicitly tailored to the landscape of the function, leading to a quasi-optimized number of hyperparameters. This allows us to  achieve high fidelity in the loaded state with high speed convergence for the studied examples.

\end{abstract}

\maketitle

\newtheorem{theorem}{Theorem}[section]
\newtheorem{corollary}{Corollary}[theorem]
\newtheorem{lemma}[theorem]{Lemma}
\def\endproof{\hfill$\blacksquare$}
%%%%%%%%%%%%%%%%%%%%%%%%%%%%%%%%%%%%%%%%%%%%%
%\section{Introduction}
%\subfile{sections/introduction}
\section{Introduction} Quantum computing has triggered a great interest in the last decades due to its theoretical capability to outperform classical information processing. Even though noise and decoherence are major drawbacks for the computational capacity of current quantum computers, quantum advantage has been experimentally achieved ~\cite{GOOGLE, PAN2, PAN}. Unfortunately, these accomplishments lack any industrial or scientific relevance, so the search of a useful application still remains. In this sense, the realistic experimental implementation of many promising quantum algorithms in several fields like solving systems of linear equations~\cite{HHL,CHILDS}, performing data fitting~\cite{FITTING}, computing scattering cross sections~\cite{SCAT1,SCAT2}, pricing financial derivatives~\cite{REBEN,WOERNER,JAVI} or initial conditions in differential equations~\cite{CHILDS2,Zanger,RIPOLL}, is constrained by the assumption that data can be efficiently loaded into a quantum device. In this context, the efficient loading of classical data into quantum computers is a particularly important problem, and represents a major bottleneck of the practical application of quantum computation in the NISQ-Era, especially with the emergence of the quantum machine learning field~\cite{MARIA1,MARIA2,ML,ML3,ML2, WIEBE}.

There exist different quantum embedding techniques transforming classical data into quantum information ~\cite{MARIA1, Havlicek}. In particular, we can distinguish two main embedding protocols depending on how the information is encoded. On the one hand, the basis embedding, in which each bit value $``0"$ or $``1"$ is mapped to a computational qubit state $|0\rangle$ or $|1\rangle$, respectively ~\cite{BRAJE}. In this way, the embedded quantum state corresponds to a uniform superposition of the bit-wise translations of binary strings. On the other hand, the amplitude-embedding technique encodes the normalized vector of classical data, which is now not necessarily binary, into the amplitudes of a quantum state~\cite{MARIA3,BLACKBOX1,BLACKBOX2,BLACKBOX3,PLESCH, MIKKO4,GROVER,KUMAR,Araujo,ZHANG,ZHAO, Bauer, Anwer}. In particular, these feature maps have been proposed to load discretized real valued functions~\cite{GROVER, MIKKO4, BLACKBOX1,BLACKBOX2} with relevant applications in loading initial conditions for solving partial derivatives equations \cite{JAVI,CHILDS2,Zanger,RIPOLL}, computing Monte-Carlo integrations \cite{WOERNER,REBEN,MONTANARO} and quantum field theory \cite{KLCO,PRESKILL}. However, a practical implementation of these approaches generally incurs into an overhead of resources, which can be reflect into either an exponential number of entangling gates~\cite{GROVER,PLESCH,MIKKO4,MIKKO1GATES,MIKKO3GATES,KUMAR}, or the employment of a huge number of ancillary qubits~\cite{Araujo,ZHANG,ZHAO}. A rather different approach sustained by the Solovay–Kitaev theorem \cite{KITAEV, NIELSEN} is based on the application of quantum generative models to efficiently accomplish an approximate amplitude encoding of discretized real valued functions~\cite{GAN,GAN2}. Nonetheless, in these generic ans\"atze, increasing the number of hyperparameters does not necessarily reflect in improving the expressability of the Ansatz to capture the function details \cite{Cerezo}. Additionally, these variational methods usually suffer from training problems such as local minima and barren plateaus \cite{McClean}.

In this article, we present two approximate quantum algorithms to load real functions into quantum computers for the NISQ era.  Our first protocol, inspired by the Grover-Rudolph algorithm and its implementation proposed by Möttönen \textit{et al} \cite{GROVER, MIKKO4},  implements the algorithm without ancillary qubits with complexity $\mathcal{O}(2^{k_0(\epsilon)})$  for functions whose second logarithm derivative is upper bounded by a constant,
with $n$ the number of qubits, $\epsilon$ the infidelity respect to the exact state, and $k_0(\epsilon)$ asymptotically independent of $n$. This leads to a dramatic reduction in the number of required two-qubit gates. Inspired by this result, we also introduce and benchmark a variational quantum circuit with applications in a broader family of functions. Our proposed Ansatz is adapted to the structure of the function, which intuitively correlates hyperparameters and expressibility. Moreover, by taking the angles provided by the Grover-Rudolph protocol, we can define a suitable initial training angle set, which considerably improves the training process, avoiding barren plateaus and local minima. Finally we have numerically proven resilience of our algorithm against several noise sources.
%The article is structured as follows. First, we briefly review the Grover-Rudolph algorithm. Then, we propose and analyze step by step our first protocol. Next, we present our variational quantum circuit and, finally, we show the results of both methods.

%%%%%%%%%%%%%%%%%%%%%%%%%%%%%%%%%%%%%%%%%%%%%
%\section{The Grover and Rudolph algorithm}\label{Section2}
%\subfile{sections/grover}

\section{The Grover-Rudolph Algorithm} 

The method of Grover and Rudolph, originally proposed in Ref. \cite{GROVER}, describes a constructive protocol to load into an $n$-qubit quantum state the discretized version $\{f_i\}$ of certain integrable density function $f\colon [x_\text{min}, x_\text{max}] \subset ${\rm I\!R}$  \to ${\rm I\!R}$^{+}$ as

\begin{equation}
\label{eq:grover}
  \ket{\Psi(f)}_n = \sum_{i=0}^{2^n-1}\sqrt{f_i}\ket{i}.
\end{equation}

\subsection{Grover and Rudolph Algorithm without ancillas}
\begin{figure}[t!]
\centering
\includegraphics[width=1\columnwidth]{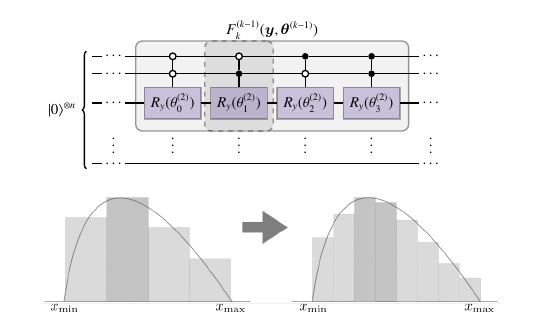}
\caption{Effect of the uniformly-controlled $y$-axis rotation $F_{k}^{(k-1)}(\bm{y}, \bm{\theta}^{(k-1)})$ for $k=3$ and $n$ qubits. Each multi-controlled gate comprising $F_{k}^{(k-1)}(\bm{y}, \bm{\theta}^{(k-1)})$ bisects the $(l+1)$-th partition interval of the function by using the conditional probability of being in the right or left side of the interval.}
\label{fig:cluster_1}
\end{figure}

The first proposal in the literature that provided an explicit circuit implementation of the Grover-Rudolph idea without using ancillary qubits was proposed by M.~ Möttönen et al \cite{MIKKO4}.  Without using ancillas, this protocol provides a constructive algorithm which applies a sequence of operation blocks, $F_{k}^{(k-1)}(\bm{y}, \bm{\theta}^{(k-1)})$, to the initial state  $\ket{0}^{\otimes n}$. For $k\in \{1,\dots,n\}$, each $k$-qubit block, $F_{k}^{(k-1)}(\bm{y}, \bm{\theta}^{(k-1)})=~ \sum_{l = 0}^{2^{k-1}-1}\ket{l}\bra{l}\otimes R_{y}(\theta_{l}^{(k-1)})$, corresponds to a uniformly-controlled $y$-axis rotation, where the $l$-th component of the angle vector $\bm{\theta}^{(k-1)}$ is calculated as
\begin{equation}
  \label{eqn:angles}
    \theta^{(k-1)}_{l}(l) = 2\arccos\left(\sqrt{\frac{\int_{x_\text{min}+ l\delta_k}^{x_\text{min} + (l+1/2)\delta_k}f(x)dx}{\int_{x_\text{min}+ l\delta_k}^{x_\text{min} + (l+1)\delta_k}f(x)dx}}\right).
\end{equation}
Here, $\delta_k = \frac{x_\text{max}-x_\text{min}}{2^{k-1}}$ and $l\in \{0, \dots, 2^{k-1}-1\}$ is the index corresponding to the $(l+1)$-th subinterval of the $2^{k-1}$ partition of the interval $[x_\text{min}, x_\text{max}]$. Each multi-controlled gate comprising $F_{k}^{(k-1)}(\bm{y}, \bm{\theta}^{(k-1)})$ is denoted by $\wedge_{k-1}\left(R_y(\theta^{(k-1)}_{l})\right)$ and bisects the $(l+1)$-th partition interval of the function by using the conditional probability of being in right or left side of the interval, as depicted in Fig. \ref{fig:cluster_1}.

In the complete algorithm, the total number of angles needed scales as $\sum_{m=1}^{n-1} 2^{m-1}=2^n-1$ , which is exponential in the number of qubits and requires an exponential number of multi-controlled C-NOT gates. Therefore, without using ancillary qubits, the required number of two qubit gates scales exponentially as the number of angles required do, and therefore the circuit complexity of implementing Grover-Rudolph algorithm without ancillas is $O(2^n)$. 

We can conclude that this protocol without ancillas is theoretically capable of loading a discretized density function at the cost of an exponential overhead of resources \cite{MIKKO1GATES,MIKKO3GATES} to prepare the state in Eq.~(\ref{eq:grover}) using blocks $F_{k}^{(k-1)}(\bm{y}, \bm{\theta}^{(k-1)})$.

\subsection{Grover and Rudolph Algorithm with ancillas}

According to the original Grover-Rudolph algorithm \cite{GROVER}, for each step $k$, we can efficiently prepare $|\Psi_k\rangle =~\sum_{l=0}^{2^k-1}\sqrt{f_l^{(k)}}|l\rangle$  by first loading the rotation angles, Eq. \ref{eqn:angles}, with bit precision $m$ into a bit string of $m$ ancillary registers and then performing $k$ controlled rotations of angle $2\pi/2^j$, $j=1,\ ...\  m$ controlled by the ancillary registers. Indeed, the encoding of the rotation angles can be efficiently achieved if we have access to the oracle:

\begin{equation}
|\Psi_k\rangle =\sum_{l=0}^{2^k-1}\sqrt{f_l^{(k)}}|l\rangle |0\rangle^{\otimes m}\rightarrow |\Psi_k\rangle =\sum_{l=0}^{2^k-1}\sqrt{f_l^{(k)}}|i\rangle \underbrace{|\theta^{(k-1)}_l(l)\rangle}_{\text{m bit precision}}
\end{equation}
%\begin{equation}
%\text{Oracle}: \sum_i |x_i\rangle |0\rangle^{\otimes k} \rightarrow \sum_i |x_i\rangle \underbrace{|\theta_i (x_i)\rangle }_{\text{k bit precision}}
%\end{equation}
An example case when this operation can be efficiently performed is when the function $\theta^{(k-1)}_l(l)$ can be well approximated by a polynomial and then, implemented by employing a polynomial amount of classical half adder operations, i.e. NAND gates. Lastly, the NAND gates are efficiently mapped into Toffoli quantum gates by making use of at most 3 qubits per classical bit \cite{CHUANG}. As this map is input-dependent, it enables us to compute it simultaneously for all $\theta^{(k-1)}_l(l)$ when the input state is a quantum superposition. However, this oracle is not explicitly provided in the original manuscript and the efficient circuit to which the authors refer is not presented, remaining as an oracle. Additionally, the original manuscript of Grover and Rudolph does not provide the analytical bounds of these approximations ($m$ bit precision, oracle implementation), as well as implicitly makes use of additional ancillary qubits which incurs into an important cost for the NISQ era. Last but not least, some recent works have risen criticism about the feasibility of this original proposal \cite{RIPOLL, PRICING, HERBET}. Moreover, to our best knowledge, there is no explicit efficient implementation of this oracle in terms of gates without employing ancillary qubits.

\begin{figure*}[t!]
  \centering
  \includegraphics[width=1.9\columnwidth]{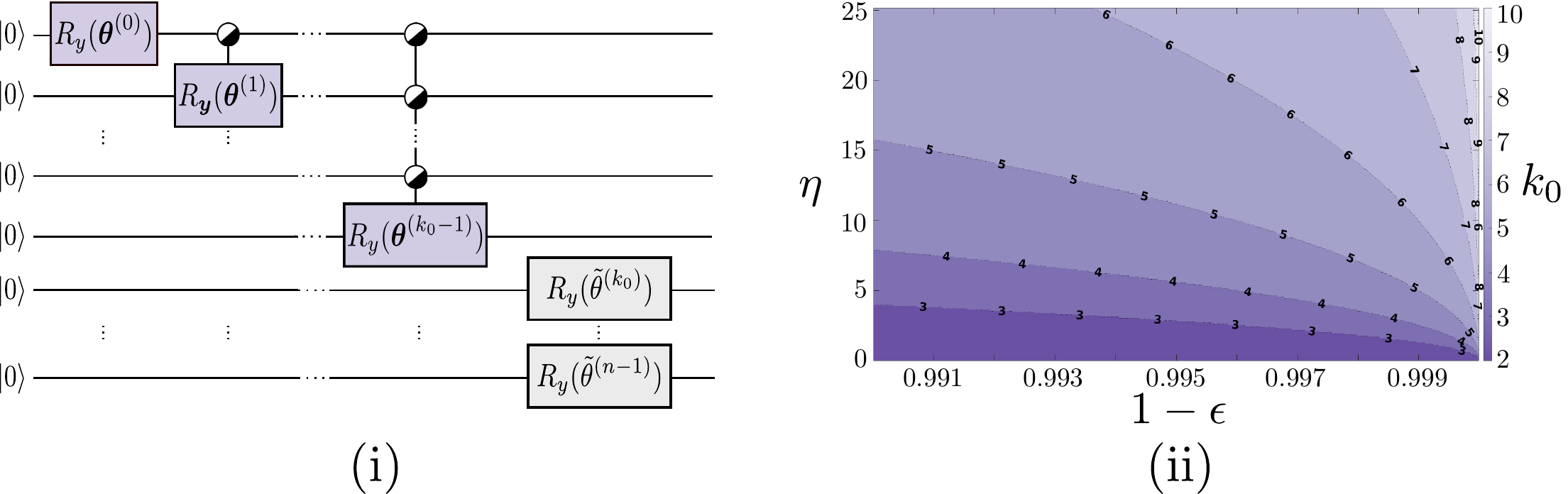}
  \caption{(i) Quantum circuit performing the protocol presented in Th.~\ref{th:k0}, based on the Grover-Rudolph method for a system of $n$ qubits. We cluster the angles of the blocks for $k\geq k_0+1>2$, leading to a drastic reduction in the number of gates needed. (ii) Values of $k_0$ for $\eta\in[0, 8\pi]$, $\epsilon\in[0.0001, 0.01]$ and $n\to\infty$, following Eq.~(\ref{eq:k0-limit}), where the dotted lines correspond to the contour lines and denote the change of values. We can appreciate that $k_0\leq 10$ for most of the cases.}
  \label{fig:circuit-heatmap}
\end{figure*}

%%%%%%%%%%%%%%%%%%%%%%%%%%%%%%%%%%%%%%%%%%%%%
%\newpage

\section{First Algorithm} Inspired by the Grover-Rudolph algorithm \cite{GROVER}, we present an efficient method to encode discretized density functions into quantum states. By permitting an error in the final state and assuming certain smoothness conditions, an angle clustering significantly reduces the required entangling gates.

\begin{defn}\label{df:f} Let $f\colon [0,1]\to ${\rm I\!R}$^{+}$ be a positive function in $L^2([0, 1])$. We define the $n$-qubit normalized representative state of $f(x)$ as the $n$-qubit state $|f(x)\rangle_n=\sum_{l=0}^{2^n-1} f(l\delta_n)|l\rangle$, with $\delta_n=\frac{1}{2^n-1}$ and $\sum_{l=0}^{2^n-1}f^2(l\delta_n) = 1$.
\end{defn}
\noindent According to this definition, we encode the discretized function into the amplitude of a quantum state, in contrast to the Grover-Rudolph algorithm, which does it in the probability,  Eq. (\ref{eq:grover}). We also consider our function defined in the interval [0, 1] as a standardization criterion.\\

\begin{thm}\label{th:k0}Let $f:[0,1]\rightarrow${\rm I\!R}$^+$ be a positive integrable function in $L^2([0,1])$ and $0\leq\eta\leq 8\pi$ a constant such that $\eta=\sup_{x\in[0,1]}\left|\partial^2_x \log f^2(x) \right|$. Then, it is posible to approximate the $n$-qubit representative state of $f(x)$, $|f(x)\rangle_n$, by a quantum state $|\Psi(f)\rangle_n$ such that the fidelity $|\langle \Psi(f)|f(x)\rangle_n|^2 \geq 1-\epsilon$ with at most $2^{k_0(\epsilon)}-1$ two-qubit gates, with
\begin{equation}\label{eq:k0}
k_0(\epsilon) = \max \left \{ \lceil -\frac{1}{2}\log_2 (4^{-n}-\frac{96}{\eta^2}\log (1-\epsilon))\rceil,2 \right \},
\end{equation}
and the circuit to perform it is provided in Fig.~\ref{fig:circuit-heatmap}\hyperref[fig:circuit-heatmap] (i).
\\\end{thm}

\begin{figure*}[t!]
  \centering
  \includegraphics[width=2.05\columnwidth]{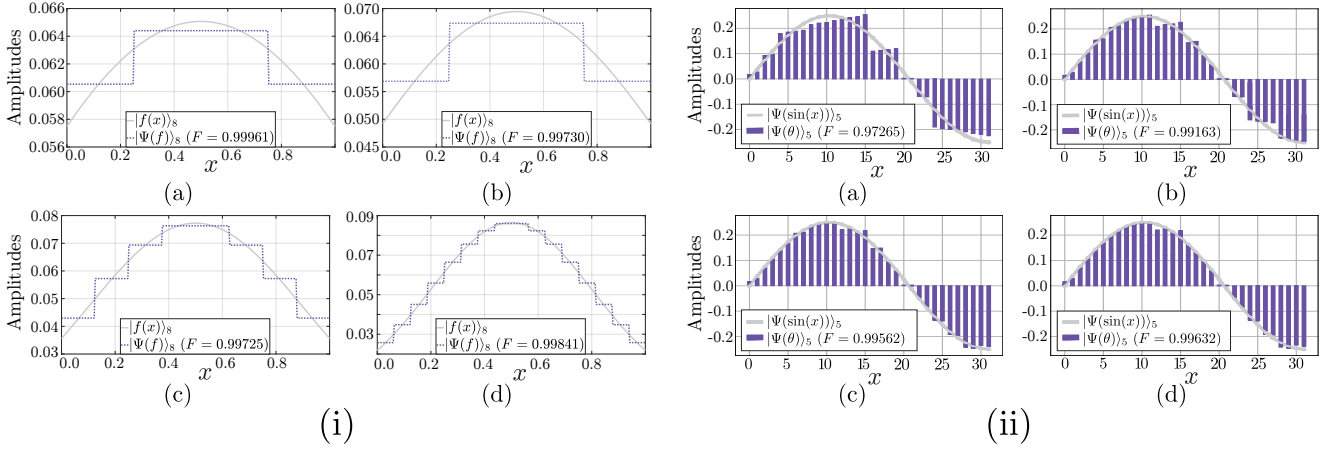}
  \caption{(i) Simulations of $\ket{f(x)}_8$ and $\ket{\Psi(f)}_8$ resulting of loading the Normal Distribution according to Th.~\ref{th:k0}, $n=8$, $\epsilon=0.05$ and different values of $\sigma$: (a) $\sigma=1$, (b) $\sigma=0.6$, (c) $\sigma=0.4$ and (d) $\sigma=0.3$. The numerical results for each experiment are given in Table~\ref{tb:normal_data}. (ii) Simulations of $\ket{\Psi\left(\sin(x)\right)}_5$ and $\ket{\Psi\left(\bm{\theta}\right)}_5$ resulting of loading the normalized sine function by training the variational Ansatz, learning rate $\gamma=1.5$, $n=5$, $k_0=2$, and different values of $p(k)$: (a) $p(k)=1$, (b) $p(k)=2$, (c) $p(k)=3$ and (d) $p(k)=k$.}
  \label{fig:results}
\end{figure*}

%%%%%%%%%%%%%%%%%%%%%%%%%%%%%%%%%%%%%%%%%%%%%%%%%%%%%%%
Let us analyze each part of the algorithm. First, we provide a sufficient condition on the target density function to guarantee an upper bound over the difference between two contiguous angles.

\begin{lem}\label{lem:2derlog} Let $f$ be a continuous function such that $f:~[0, 1]\to${\rm I\!R}$^+$ and consider a block comprising a uniformly-controlled rotation of $k$ qubits. Then, the difference between two contiguous angles is bounded by the second derivative of the logarithm of $f$ in the following way:
\begin{equation}
  \left|\theta^{(k-1)}_{l+1}-\theta^{(k-1)}_l\right| \leq \frac{\delta_k^2}{4}\max_{y'\in[l\delta_k, (l+1)\delta_k]}\left|\left.\left(\partial^2_{y}\log f(y)\right)\right\rvert_{y=y'}\right|,
  \label{eq4}
\end{equation}
where $\delta_k=\frac{1}{2^{k-1}}$, $l\in\{0, \dots, 2^{k-1}-2\}$.
\end{lem}
Consequently, if $\left|\partial^2_{y}\log f(y)\right| \leq \eta$, then for each block comprising a uniformly-controlled rotation of $k$ qubits, the difference between any two angles is bounded by

\begin{equation}
\label{eq:cluster}
    \left|\theta^{(k-1)}_{l}-\theta^{(k-1)}_{l'}\right| \leq \frac{\delta_k}{4}\eta \ \ \ \   \forall l,l' \in \{0, \dots, 2^k-1\}.
\end{equation}
This result allows us to cluster angles of each block in which Eq.~(\ref{eq:cluster}) is fulfilled. Thus, we define a cluster representative angle, $\tilde{\theta}^{(k-1)}$,  as
\begin{equation}
\label{eq:representative}
    \left|\tilde{\theta}^{(k-1)}-\theta^{(k-1)}_{l}\right| \leq \frac{\delta_k}{8}\eta\vcentcolon=\eta_k \ \ \ \   \forall l \in \{0, \dots, 2^k-1\},
\end{equation}
\noindent $\forall k\geq k_0$, with $k_0+1$ the index of the first block in which Eq.~(\ref{eq:cluster}) is fulfilled (successive blocks also verify it). Note that $\left|\partial^2_{y}\log f(y)\right| \leq \eta$ is a sufficient condition for the bound in Eq.~(\ref{eq4}), however, if in a singularity point it grew slower than $\frac{1}{\delta^2_k}$, the difference between consecutive angles still vanishes as we will analyze later.
%%%%%%%%%%%%%%%%%%%%%%%%%%%%%%%%%%%%%%%%%%%%%%%%%%%%%%%

We now analyze how clustering the angles according to Eq.~(\ref{eq:representative}) affects the final error of the process, measured by means of the fidelity with respect to the exact discretized state $|f(x)\rangle_n$. Considering an $n$-qubit system, the unitary gate to prepare the quantum state representing the target density function can be written in terms of the blocks as $\mathfrak{U}_n =~\mathcal{U}_{n-1}\left(\bm{\theta}^{(n-1)}\right)\ldots\mathcal{U}_{0}\left(\bm{\theta}^{(0)}\right)$, where we define $\mathcal{U}_{k-1}\left(\bm{\theta}^{(k-1)}\right) \coloneqq~ F_{k}^{k-1}(\bm{y}, \bm{\theta}^{(k-1)})\otimes\iden{(n-k)}.$

Let $\tilde{\mathfrak{U}}_n$ denote the operation $\mathfrak{U}_n$ when the rotation angles corresponding to each block $k$ are replaced by a representative, $\tilde{\theta}^{(k-1)}$, such that its difference with any angle of the block is at most $\eta_k$, i.e. $|\theta_l^{(k-1)} -\tilde{\theta}^{(k-1)}|\leq \eta_k$ for $l=0, \dots, 2^{k-1}-1$ and $k=1, \dots, n$. Then, the following lemma can be proven:
\\
\\

\begin{lem} Consider a system of $n$ qubits and an error $\eta_k$ between any angle of the $k$-th block and its representative such that $\eta_k\leq\pi$, with $k=1, \dots, n$. Then, the fidelity between the final states with and without clustering, $F=~|\brazero{n}\mathfrak{U}_n^\dagger \tilde{\mathfrak{U}}_n\ketzero{n}|^2$, satisfies
  \begin{equation}
  F \geq \prod_{k=1}^n\cos^{2}\left(\eta_k/2\right).
  \end{equation}
\end{lem}

Assuming that we cluster angles of the blocks for $k\geq k_0+1>2$, then
\begin{equation}
F \geq \prod_{k=k_0+1}^n\cos^{2}\left(\frac{\eta}{8 2^k}\right)\geq e^{-\frac{\eta^2}{96}(4^{-k_0}-4^{-n})}\vcentcolon= F_{k_0},
\end{equation}
since $\cos(x) \geq e^{-x^2}$ for $x\lessapprox\pi/2$. Therefore, if an infidelity $\epsilon=1-F_{k_0}$ is allowed and the angles of all blocks comprised of more than $k_0(\epsilon)$ qubits are clustered, with $k_0$ given by Eq.~(\ref{eq:k0}), then the fidelity satisfies $F\geq 1-\epsilon$. In the asymptotic limit of $n\to\infty$, we have that $k_0(\epsilon)$ tends to
\begin{equation}\label{eq:k0-limit}
  k_0(\epsilon) \to \max \left \{ \lceil -\frac{1}{2}\log_2 (-\frac{96}{\eta^2}\log (1-\epsilon))\rceil,2 \right \},
\end{equation}
which is independent of the system size, $n$. In Fig. (\ref{fig:circuit-heatmap})\hyperref[fig:results] (ii), we have depicted the values of $k_0$ at this limit, for $\eta\in[0, 8\pi]$ and $\epsilon\in[0.0001, 0.01]$.
\\
\\
We finally study the implementation cost of the proposed protocol. We only take into account the latter type of gates and ignore single-qubit operations \cite{2qubit-efficient}. Using the result of \cite{MIKKO1GATES,MIKKO3GATES}, which illustrates how a uniformly-controlled rotation of $k$ qubits can be implemented with $2^{k-1}$ CNOTs, the complexity of the circuit described in Th.~\ref{th:k0} is $\mathcal{O}(2^{k_0(\epsilon)})$.% instead of the $\mathcal{O}(2^n)$ required by the Grover-Rudolph algorithm.

\subsection{Normal Distribution} We apply the algorithm given by Th.~\ref{th:k0} to a normal distribution with a mean value $\mu=0.5$ and for different values of the variance $\sigma$. We numerically benchmark the fidelity attained by our first protocol using the value of $k_0$ resulting of Eq.(\ref{eq:k0}) when we assume a maximum infidelity of $\epsilon=0.05$ and a system of $8$ qubits.  Notice that for this distribution, $\eta=2/\sigma^2$. In Fig.~\ref{fig:results}\hyperref[fig:results] (i) and Table~\ref{tb:normal_data}, we have depicted the results from the simulations of $\ket{f(x)}_8$ and $\ket{\Psi(f)}_8$, for different values of $\sigma$. We appreciate that the condition of $F\geq0.95=1-\epsilon$ is not only satisfied in all cases, but the fidelities obtained are considerably better. Furthermore, the significant reduction in the number of two-qubit gates required to achieve these results is noteworthy. In the worst case, for $\sigma=0.3$, the quantity of gates needed represents the $12.16\%$ of the original set, while the fidelity of the experiment reaches $0.99841$.

\begin{table}[b!]
\caption{Numerical data for the simulations depicted in Fig.~\ref{fig:results}\hyperref[fig:results] (i). The values of $k_0$ have been computed using Eq. (\ref{eq:k0}). In addition, the number of two-qubit gates is given by $2^{k_0}-1$, and the last column is the percentage between the required gates and the total given by the Grover-Rudolph algorithm, which for $n=8$ are 255.}

\begin{ruledtabular}
\begin{tabular}{cccccc}
 $\sigma$&$\eta$&$k_0$&Fidelity&\#TQG&\%TQG\\
\hline
1.0 & 2.00  & 2 & 0.99961 & 3  & 1.18  \\
0.6 & 5.56  & 2 & 0.99730 & 7  & 2.75  \\
0.4 & 12.50 & 3 & 0.99725 & 15 & 5.88  \\
0.3 & 22.22 & 4 & 0.99841 & 31 & 12.16 \\
\end{tabular}
\end{ruledtabular}\label{tb:normal_data}

\end{table}

\subsection{Generalization: Singular Points}\label{ap:general}
In this section we study the generalization of the Th. \ref{th:k0} when the functions are allowed to have singularities on the boundary that grow slower than the size of the grid.

%\subsection{Singularities in $\left|\partial^2_x \log f^2(x) \right|$ }\label{ap:sing}
%\subfile{sections/appendix/singularities}
Consider a function $f:[0,1]\rightarrow\mathbb{R}^+$ with
\begin{equation}
\left|\partial^2_{x}\log f(x)\right|\xrightarrow[x \to 0]{} \infty.
\end{equation}
An example is the beta density function, defined as
\begin{equation}
  f(x)=\frac{x^{\alpha-1}(1-x)^{\beta-1}}{B(\alpha, \beta)},
\end{equation}
with $\alpha, \beta >0$ and $B(\alpha, \beta)$ the beta function. Then, the second derivative of its logarithm is
\begin{equation}
  \partial^2_{x}\log f(x)=\frac{1-\alpha}{x^2}+\frac{1-\beta}{(1-x)^2}.
\end{equation}
Thus, for $\alpha\neq 1$, we have a singularity at $x=0$. Also, we see that if $\beta\neq 1$, there is a singularity at $x=1$.

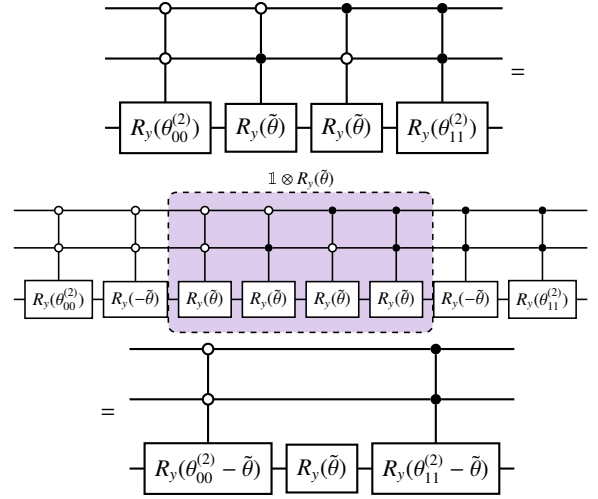
\begin{figure}[t!]
\begin{quantikz}[column sep=0.2cm]
& \octrl{1}  & \octrl{1} & \ctrl{1} & \ctrl{1}& \qw \\
& \octrl{1}  & \ctrl{1} & \octrl{1} & \ctrl{1}& \qw \\
& \gate{R_y(\theta^{(2)}_{00})} & \gate{R_y(\tilde{\theta})} & \gate{R_y(\tilde{\theta})} & \gate{R_y(\theta^{(2)}_{11})} & \qw
\end{quantikz}
%\newline
=
\begin{adjustbox}{width=.9\columnwidth}
\begin{quantikz}[column sep=0.2cm]
& \octrl{1} & \octrl{1} & \octrl{1}\gategroup[3,steps=4,style={dashed,rounded corners,fill=violet!70!blue!20!, inner xsep=2pt},background]{$\mathbbm{1}\otimes R_y(\tilde{\theta})$}  & \octrl{1} & \ctrl{1} & \ctrl{1} & \ctrl{1} & \ctrl{1}& \qw \\
& \octrl{1} & \octrl{1} & \octrl{1}  & \ctrl{1} & \octrl{1} & \ctrl{1} & \ctrl{1} & \ctrl{1}& \qw \\
& \gate{R_y(\theta^{(2)}_{00})} & \gate{R_y(-\tilde{\theta})} & \gate{R_y(\tilde{\theta})} & \gate{R_y(\tilde{\theta})} & \gate{R_y(\tilde{\theta})} & \gate{R_y(\tilde{\theta})} & \gate{R_y(-\tilde{\theta})} & \gate{R_y(\theta^{(2)}_{11})} & \qw
\end{quantikz}
\end{adjustbox}
\newline
=
\begin{quantikz}[column sep=0.2cm]
& \octrl{1}  & \qw & \control{} \vqw{1}& \qw \\
& \octrl{1}  & \qw & \ctrl{1}& \qw \\
& \gate{R_y(\theta^{(2)}_{00}-\tilde{\theta})} & \gate{R_y(\tilde{\theta})} & \gate{R_y(\theta^{(2)}_{11}-\tilde{\theta})} & \qw
\end{quantikz}
\caption{Example of the reduction of gates for the block of three qubits $F_{3}^2(\bm{\hat{y}}, \bm{\theta}^{(2)})$, where it is assumed that both $\theta^{(2)}_{01}$ and $\theta^{(2)}_{10}$ can be approximated by $\tilde{\theta}$.}
\label{fig:cluster}
\end{figure}

 In this situation, $\left|\partial^2_{x}\log f(x)\right|$ can not be bounded by a finite factor $\eta$ in the whole interval and, hence, Th.~\ref{th:k0} can not be applied. However, under certain circumstances, this issue can be solved. Recall that, in Lem.~\ref{lem:2derlog}, we obtained that the difference between two consecutive angles in a block of $k$ qubits satisfies
\begin{equation}
  \left|\theta^{(k-1)}_{l+1}-\theta^{(k-1)}_l\right| \leq \frac{\delta_k^2}{4}\left|\partial^2_{x}\log f(x)\right|,
\end{equation}
with $\delta_k=\frac{1}{2^{k-1}}$, $l\in\{0, 1, \dots, 2^{k-1}-2\}$, and $x=l\delta_k$. Since the singularity is found in $x=0$, there exists a $k_{\text{max}}$ from which the maximum of $\left|\partial^2_{x}\log f(x)\right|$ in $[2^{-k_{\text{max}}+1}, 1]$ is found in $x=2^{-k_{\text{max}}+1}$. Then,
\begin{equation}
  \left|\theta^{(k_{\text{max}}-1)}_{l+1}-\theta^{(k_{\text{max}}-1)}_l\right| \leq \frac{1}{4}\frac{\left|\partial^2_{x}\log f(x)|_{x=2^{-k_{\text{max}}+1}}\right|}{2^{2k_{\text{max}}-2}}.
\end{equation}
Next, if for the limit $k\to\infty$ the term $\left|\partial^2_{x}\log f(x)|_{x=2^{-k+1}}\right|~2^{-2k+2}\to 0$, this value is decreasing and there exists a $k^*\geq~ k_{\text{max}}$ from which an $\eta=\left|\partial^2_{x}\log f(x)|_{x=2^{-k^*+1}}\right|$ can be set so the clustering conditions $\frac{\delta_k}{8}\eta \leq \pi \implies \eta \leq 2^{k-1}8\pi$ are satisfied. This $k^*$ acts as the $k_0$ of Th.~\ref{th:k0} and represents the last block without clustering. Then, for $k>k^*$, the angles corresponding to the interval $[2^{-k+1}, 1]$ can be clustered, following our protocol. The case for a singularity in $x=1$ is analogous and in Fig.~\ref{fig:cluster}  the resulting gates corresponding to the process of clustering the inner angles for a block of 3 qubits are depicted. Given that $R_y(\tilde{\theta}) R_y(-\tilde{\theta})=\mathbbm{1}$, we can complete the identity of the clusterized block by subtracting $\tilde{\theta}$ from the rest of the angles.

 In this situation, if we consider that singularities exist in $x=\{0, 1\}$, the total number of two-qubit gates required to load $f$ into a quantum state is
 \begin{align}
   \# \text{TQGs} &= \sum_{k=1}^{k^*}2^{k-1}+2\sum_{k=k^*+1}^{n}(80k-398)\nonumber\\&=2^{k^*}-1+2(n-k^*)(20k^*+20n-179),
 \end{align}
 see the Appendix A for further details in gates decomposition. Notice that this result is only valid for $k^*>6$, but since we are interested in the asymptotical behaviour of the protocol, it is not an issue. All in all, we obtain that the complexity of this process is exponentially dependent on $k^*$ with an extra polynomial term.

 On the other hand, if a singularity is found in $(0, 1)$, we end up with multiple clusters of angles in each block. The reason behind this is that the clusters are formed with contiguous angles. Therefore, the reduction of gates is not significant, and the protocol can not be performed efficiently (polynomial). However, if the representatives of the disjointed clusters are equal, then we can create a single cluster, so the reduction is doable. In this sense, we have numerically observed that far from the singularities, all angles converge to a value of $\pi/2$, but we have not found an analytical proof yet.

 Let us see an example of a function that meets the previous description and analyze the outcome of the protocol. Consider the function
 \begin{equation}\label{eq:poli}
   f(x)=\frac{1}{N}e^{x^{3/2}}
 \end{equation}
 in $[0, 1]$, where $N$ is the normalization factor, and a system of $n=10$ qubits. The second derivative of its logarithm is
 \begin{align}
   \log f(x) = x^{\frac{3}{2}} -\log N
   &\implies \partial_{x}\log f(x) = \frac{3}{2}\sqrt{x}\\
   &\implies \partial^2_{x}\log f(x) = \frac{3}{4\sqrt{x}}.
 \end{align}
 Hence, $\partial^2_{x}\log f(x)$ has a singularity at $x=0$.
First of all, since $\partial^2_{x}\log f(x)$ is a monotonic decreasing function, its maximum in the interval $[2^{-k+1}, 1]$ is found in $x=2^{-k+1}$, for any $k=~1, \dots, n$. Then, we can set $k_{\text{max}}=1$. Next, we need to compute the limit of $\left|\partial^2_{x}\log f(x)|_{x=2^{-k+1}}\right|2^{-2k+2}$:
 \begin{align}
   &\left|\partial^2_{x}\log f(x)|_{x=2^{-k+1}}\right|2^{-2k+2} = \frac{3}{4\sqrt{x}}|_{x=2^{-k+1}}2^{-2k+2}\nonumber\\
   &\quad=\frac{3}{4} 2^{\frac{1}{2}(k-1)}2^{-2k+2}=\frac{3}{4} 2^{-\frac{3}{2}k+\frac{3}{2}}\xrightarrow[k \to \infty]{} 0.
 \end{align}
 Therefore, the difference in the angles is bounded and we can select a $k^*\geq 1$ for which the conditions of Th. $\ref{th:k0}$ are satisfied in $[2^{-k^*+1}, 1]$. The first condition we need to check is the inequality given by
 \begin{equation}
   \frac{\delta_k}{8}\eta \leq \pi,
 \end{equation}
with $\eta=\left|\partial^2_{x}\log f(x)|_{x=2^{-k+1}}\right|=\frac{3}{4}2^{\frac{1}{2}(k-1)}$. Then,
\begin{align}
  \frac{3}{32}2^{\frac{1}{2}(k-1)}2^{-k+1}=\frac{3}{32}2^{\frac{1}{2}(1-k)}.
\end{align}
Since this term is decreasing, its maximum is found when $k=1$, with a value of $\frac{3}{32}<\pi$. Thus, this condition is met for any $k=1, \dots, n$, so we can select $k^*=1$. Additionally, if we compute the value of $k_0$ with $\epsilon=0.01$ and $\eta=\left|\partial^2_{x}\log f(x)|_{x=2^{-k^*+1}}\right|=0.75$ using Eq.~(\ref{eq:k0}), we obtain $k_0=2$. Then, the last block to remain unclustered must be the maximum between $k_0$ and $k^*$, which in this case is 2.

\begin{figure}[t!]
   \centering
   \includegraphics[width=1\columnwidth]{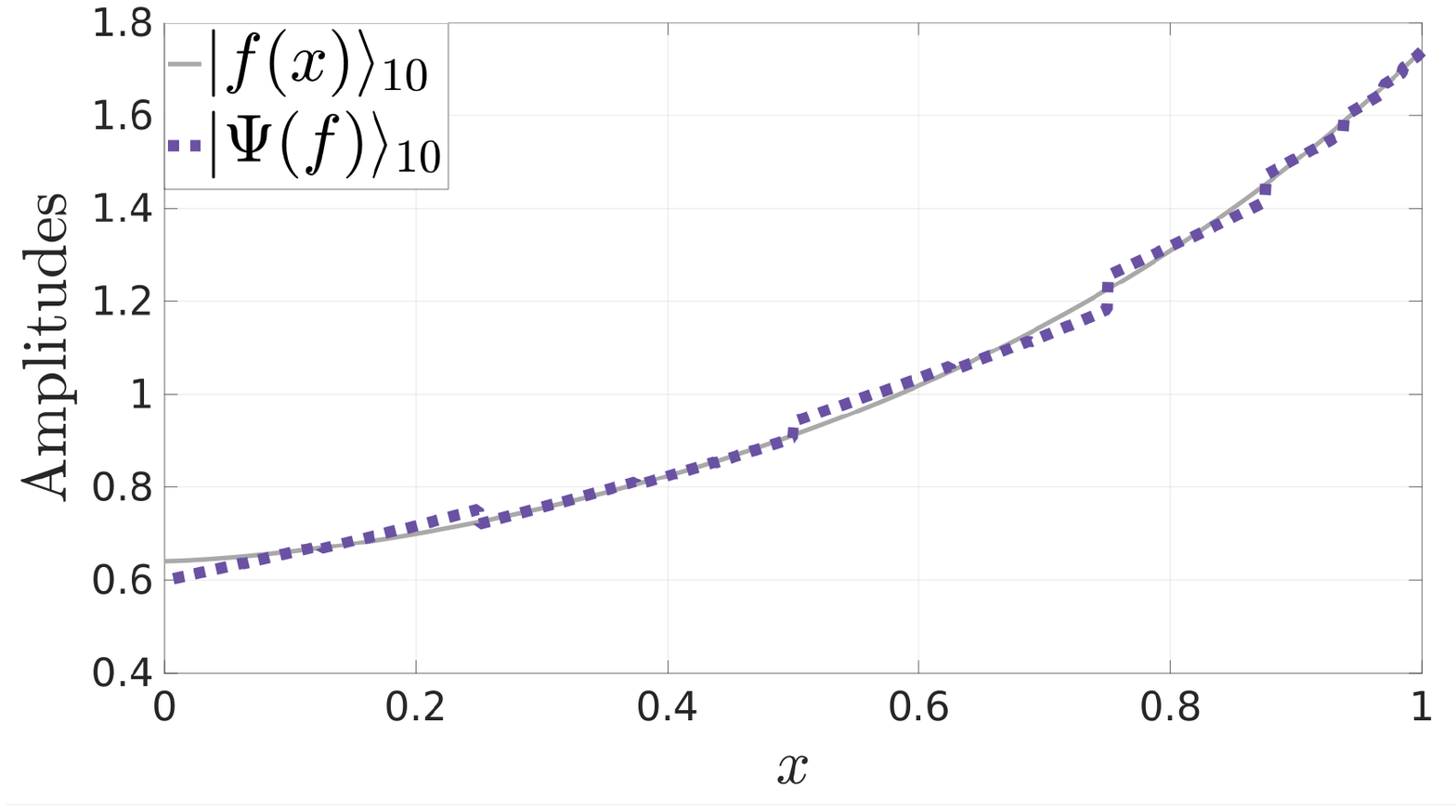}
   \caption{Simulation of $\ket{f(x)}_{10}$ and $\ket{\Psi(f)}_{10}$ for the function defined in Eq.~(\ref{eq:poli}), $n=10$, $\epsilon=0.01$ and the clustering process performed for $k=3, \dots, n$.}
   \label{fig:poli}
 \end{figure}

 Now, in Fig.~\ref{fig:poli}, we have depicted the result of the experiment of the considered function for a system of $n=10$ qubits and the clustering starting with the 3-qubit block, following the protocol described in this section. With a fidelity of $0.99975$, larger than the one required, we have that the final state $\ket{\Psi(f)}_{10}$ successfully captures the features of $f$ with a reduction of the complexity of two-qubit gates from $\mathcal{O}(2^{10})$ to $\mathcal{O}(2^{2})$.

 \subsection{Analysis of resilience to noise in \textit{NISQ} era}

\begin{figure}[b!]
\includegraphics[width=1.02\columnwidth]{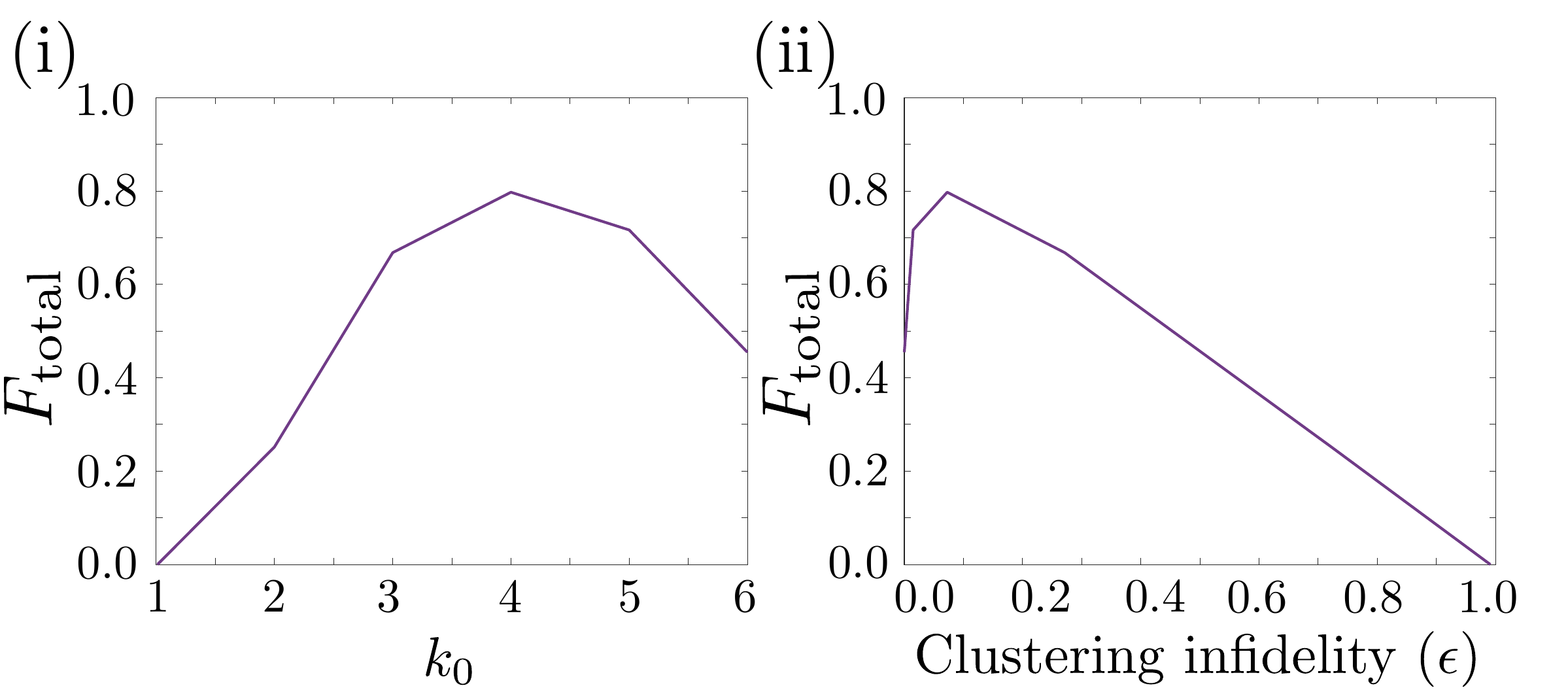}
\caption{Total fidelity, $F_{total}$, in terms of the clustering infidelity assumed in the protocol according to the expression $\epsilon(k_0) \sim 1-e^{-\eta^2/24(4^{-k_0} - 4^{-n})}$, where $\eta=22.22$, $k_0=1,...,6$  and $n=6$ . As we can appreciate, the expected fidelity of our protocol is above the fidelity resulting of implementing the protocol without clustering in noisy devices, which dramatically tends to 0. Actually, the fidelity reaches a maximum for $k_0=4$ , which corresponds to a clustering error $\epsilon_0=0.0726.$ }
\label{fig:reply_1}
\end{figure}

In this subsection, we present a theoretical and numerical analysis of how experimental errors affect different clustering levels in our first protocol and their impact on the final fidelity. The crucial point here is that when digital accuracy increases, the number of gates requested grows exponentially. As the introduction of these gates implies a growing experimental error in \textit{NISQ} quantum processors, we observe a trade-off, see Fig. \ref{fig:reply_1}, between the clusterization error and the experimental error, quite similar to the trade-off observed in digital quantum simulations between the number of Trotter steps and the experimental error in Refs.~\cite{Urtzi, Urtzi2}. This balance is crucial for algorithms in noisy quantum processors. Reproducing a similar reasoning as the one in the aforementioned reference and references thereof, we first propose an approximated model of how the experimental error combined with the clustering error of our algorithm scales as a function of the number of non-parallel two-qubit gates. Then, we perform some numerical simulations introducing multiple realistic noises in our algorithm to support our theoretical predictions.

In order to establish a theoretical framework to understand the behavior of our system when clusterization  and experimental errors are considered, we make the assumption that the main source of experimental noise comes from the two-qubit gates, while ignoring the noise arising from single-qubit rotations. Additionally, we consider that the application of each of these gates onto a quantum state, denoted as $\rho$, is modeled in the following form

\begin{equation}
\rho \rightarrow (1-\xi) U^\dagger \rho U + \xi \tilde{U}^\dagger \rho \tilde{U}
\end{equation}

\begin{figure*}
\includegraphics[width=2.05\columnwidth]{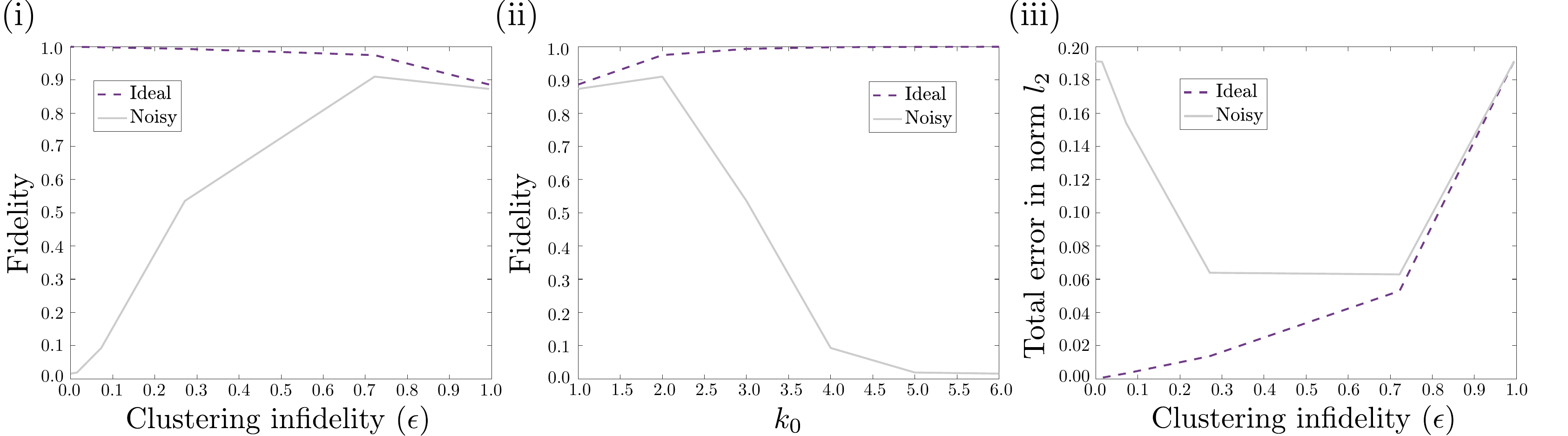}

\caption{(i) and (ii) Total fidelity, $F_{total}$, in terms of the clustering infidelity assumed and $k_0$  in the protocol according to the expression $\epsilon(k_0) \sim e^{-\eta^2/24(4^{-k_0} - 4^{-n})}$, where $\eta=22.22$,  $k_0=1,...,6$ and $n=6$  for the ideal and noisy cases. (iii) $l_2$ normalized error in terms of the clustering in the same conditions. The maximal fidelity, respectively the minimal error measure with the $l_2$ norm, is achieved for a clusterization level $k_0=2$ , which corresponds to  $\epsilon_0=0.7222$}
\label{fig:reply_2}
\end{figure*}

Where $U$  corresponds to the desired dynamics while the second term introduces a certain Taylor expansion of the deviation from the exact evolution, with $\xi \ll 1$. After $m$ non-parallel gates, ignoring quadratic terms in $\xi$, the infidelity of the state has approximately evolved according to

\begin{equation}
1-F_{\text{exp}}\sim m\ \xi \ \|\tilde{U}\|
\end{equation}
as the leading term. Note that the argument could be extended by replacing $\tilde{U} \rightarrow T(\cdot)$ , an arbitrary quantum channel, but the calculation is more complicated to produce a similar argument. In a rough approximation, for a certain value of the clustering index $k_0$, the total fidelity of the protocol run on a \textit{NISQ} device has a contribution coming from this experimental noise together with the digital infidelity due to the clustering procedure. We can approximate this quantity by

\begin{equation}
F_{\text{total}}\sim e^{-\eta^2/24(4^{-k_0} - 4^{-n})}-2\xi\|\tilde{U}\| (-1 + 2^{k_0}) 
\end{equation}
which in terms of the clustering error, $\epsilon$ and by using that $1-\epsilon \sim e^{-\eta^2/24(4^{-k_0} - 4^{-n})}$, and hence,  $2^{k_0}-~1 \sim~ \left ( 4^{-n}-\frac{96}{\eta^2}\log (1-\epsilon)\right)^{-\frac{1}{2}}-~1$,  can be expressed as

\begin{equation}
F_{\text{total}}\sim 1 - \epsilon -2\xi\|\tilde{U}\| \left (\left ( 4^{-n}-\frac{96}{\eta^2}\log (1-\epsilon)\right)^{-\frac{1}{2}}-1\right).
\end{equation}

From the equation above, we can find the value of clustering error $\epsilon_0$  for which  $F_\text{total}$ achieves its maximum by deriving the function. This condition holds

\begin{equation}
4\,(1-\epsilon_0)^2 \, \left (4^{-n} -\mu \log (1-\epsilon_0)\right)^{3} = \alpha^2 \mu^2
\end{equation}
with $\alpha = 2\xi\|\tilde{U}\|$  and $\mu = 96/\eta^2$ , which is a transcendent equality, so it cannot be analytically solved. However, for the range of parameters in which we are interested and for  sufficiently small, we can approximate $\epsilon_0 \sim \frac{\alpha \,2^{n-\frac{3}{2}}}{\sqrt 3}$.

In order to provide a numerical example for this expression, we analyze a normal distribution with  $\sigma=0.3$ encoded in  $n=6$ qubits. For the experimental error, we consider $\alpha=~2\xi||\tilde{U}|| = 0.0087$. We depict the results of this analysis in Fig. 1. If we compare results from this analysis, we can see that the maximum fidelity is achieved for a value of $\epsilon_0=0.0726$ while the predicted value reads $\epsilon_0=0.1131$.

Once the theoretical framework has been established, let us carry out some numerical simulations analyzing the robustness (in terms of fidelity) against different noises. The main objective is to support the aforementioned analytical findings. The guidelines of the numerical experiment is described as follows. The circuit is transpiled to a native set of gates given by \textit{CNOT, Id, Rz $(\theta)$, X} and  \textit{Sx}. The noise quantum channels considered are
\begin{itemize}
\item Bit-Flip (Pbf)
\item Amplitude-Damping (T1)
\item Dephasing (T2)
\item Gate errors (rD, CNOT error)
\item Measurement error (pmeas)
\end{itemize}

\begin{table}[b!]
\centering
\begin{tabular}{c||c||c}
\hline
\hline
\textbf{Parameter} & \textbf{Description} & \textbf{Value} \\
\hline
\hline
SQG time & Single qubit gate time (ns) & 35 \\
CX time & CX gate time (ns) & 540 \\
rD & Deviation ratio for the single qubit gates & 2,457E-04 \\
Pbf & Bit-flip error during the rz gate & 2,457E-04 \\
CNOTerror & Deviation ratio for CX gate & 8,328E-03 \\
pmeas & Readout error & 2,23E-01 \\
pth & Thermal population of the ground state & 0.01 \\
T1 & Decoherence time (us) & 114.84 \\
T2 & Dephasing time (us) & 38.65 \\
\hline
\end{tabular}
\caption{Noise parameters description and their value. We have estimated the numerical values from the calibration data provided for the IBM device `ibm\_jakarta'.}
\label{tab:features}
\end{table}

\begin{figure*}[ht!]
  \centering
  \includegraphics[width=2\columnwidth]{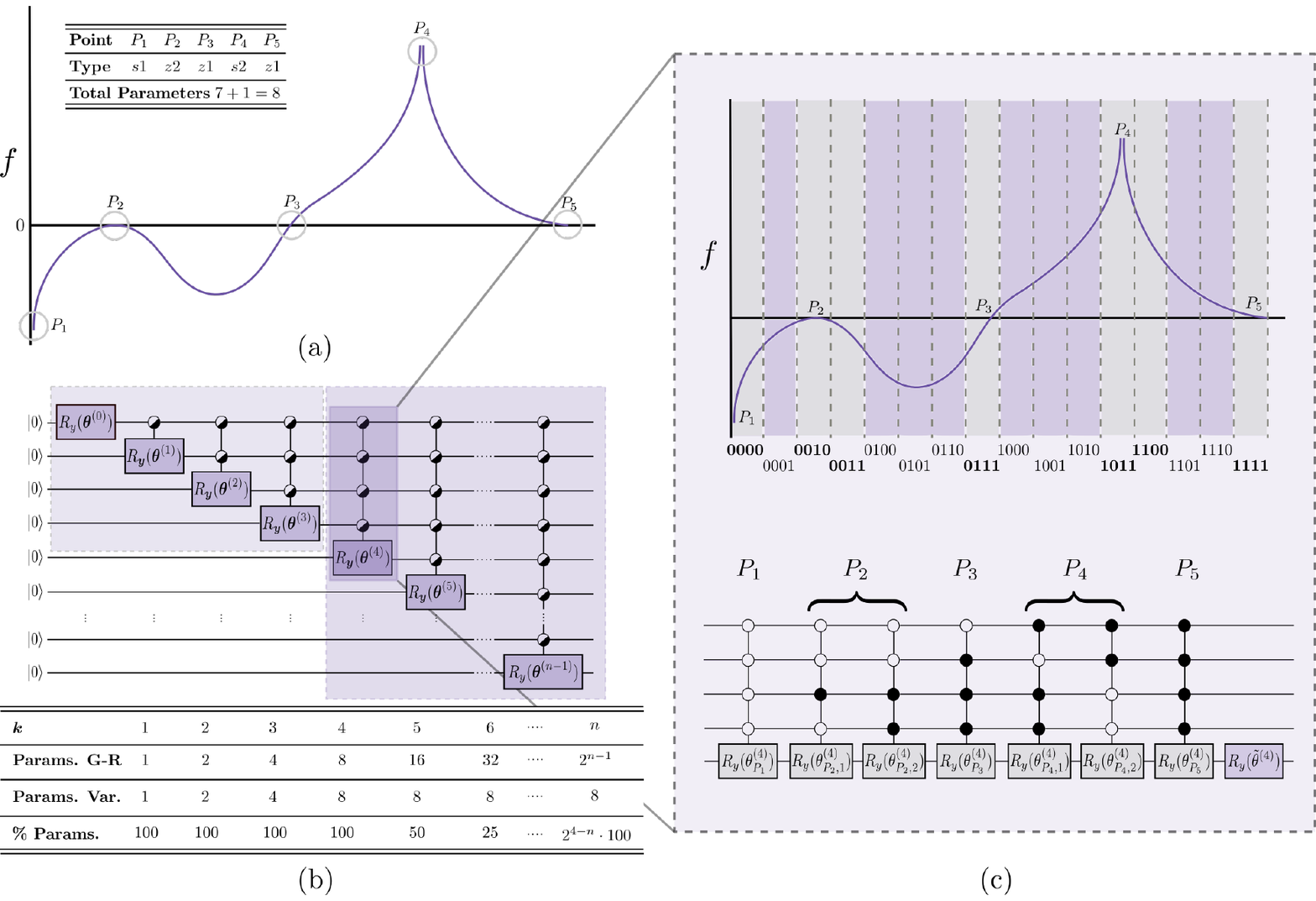}
  \caption{Illustration of proposed variational circuit for a general function. (a) Target function. The proposed landscape generalizes the possible cases of special points that lead to singularities in the second logarithmic derivative. The special points are classified in the legend of the graph, where $z$ denotes the zeros, $s$ the singularities, and the index denotes whether the slope in the contiguous points has the same sign, index$=1$, or not, index$=2$. This is translated to a minimum amount of 7 necessary parameters required to capture the local behavior of the function in these points, one per index$=1$ and 2 per index$=2$. We add an additional parameter which represents the clusterization. (b) Variational circuit. The blocks which have a number of angles lower or equal to the necessary parameters will remain unclustered. From the first block which has more angles, we proceed to reduce the number of angles to the number of parameters by clustering the ones that do not correspond to the special points. This leads to a significant reduction in the number of gates. (c) Zoom in of the  $F_5^{4}(y, \bm{\theta})$ block clusterization. As we only need $8$ parameters, the $16$ angles of this block are reduced to the half. We keep one local rotation for the clustering parameter and 7 multi-controlled gates corresponding to the intervals where the singularity is located.}
  \label{fig:general}
\end{figure*}

For a realistic scenario, we have taken the value of these errors from the IBM Jakarta quantum processor, which are summarized in Table \ref{tab:features}. Considering this set up, we have studied again the normal distribution with $\sigma=0.3$ encoded in $n=6$ qubits, focusing on the fidelity and  $l_2$ norm (normalized with the system size such that it converges to the  $L_2$ norm in the continuous limit) with respect to the exact discretized state. The numerical results are depicted in Fig. \ref{fig:reply_2}, which also shows a trade-off between the clustering and the experimental error. More explicitly, the maximal fidelity, respectively the minimal error measure with the  norm $L_2$, is achieved for a clusterization level $k_0=2$ , which reproduces the structure showing a trade off between errors predicted by our theoretical model. However the maximum is reached for a smaller value of $k_0$ , compared to the predictions, as the theoretical model is a first order simplification which becomes not that accurate when the presence of more kind of noises is assumed. This means that our idea based on clustering can be implemented with shallow but not trivial circuits and presumably offers a robust performance for the \textit{NISQ} era.

Consequently, although our protocol introduces a controllable error, it significantly reduces the depth required for the computation as well, resulting in a more balanced and reliable outcome. In the presence of experimental noise, our algorithm achieves a good balance between fidelity (experimental + clustering errors) and feasibility (realistic depth), which is crucial for this stage of quantum computing, where computers are characterized by high error rates and limited coherence times.

\section{Variational Quantum Circuit}
\subsection{Ansatz}

Based on the previous protocol, we propose a variational Ansatz for loading functions beyond the conditions required by Th. 1. We consider  $(z+s+1)p(k)$ hyperparameters for each $k$-qubit block satisfying $k>k_0$,  where $z$ and $s$ are respectively the number of zeros and singular points in the function, $p(k)$ is polynomial number in $k$ denoting the number of hyperparameters allowed per singular point/zero for the $k$-th block, and $k_0=\text{max}\{\ k \  | \ z+s+1\geq 2^k\}$. Assuming $p(k)\geq1$, the minimum number of hyperparameters needed to capture the singular behavior of the function is $z+s+1$. Therefore, for each $k$-qubit block $F_{k}^{(k-1)}(\bm{\hat{y}}, \bm{\theta})$ comprising more than $z+s+1$ parameters, we cluster the angles which do not correspond to the position of zeros or singularities.
In this form, this proposal establishes an intuitive correlation between the number of hyperparameters $p(k)$ and the expressability of the circuit to capture the details of the target functions in the most relevant points. By using the decomposition of the multi-controlled rotations \cite{MIKKO1GATES,MIKKO3GATES}, the number of two-qubit gates which must be included in the variational circuit sums up to
\begin{equation*}
  \# \text{TQGs} = \sum_{k=1}^{k_0}2^{k-1}+(s+z)\sum_{k=k_0+1}^{n}p(k)(80k-398)
\end{equation*}
\begin{equation}
  =2^{k_0}-1+r_{k_0+1}(n, s+z),
\end{equation}
with $m$  the number of zeros/singularities and $r_{k_0+1}(n, s+z)$ growing polynomially with the system size $n$. A remarkable advantage of this Ansatz is the training procedure, since Grover-Rudolph algorithm provides a suitable set of initial training angles which considerably enhances the convergence of the protocol with respect to a random initialization. This fact, together with a scaling in the number of hyperparameters $p(k)$ substantially slower than the system size, allow us to avoid the training procedure to get stuck into both local minima and barren plateaus. In Fig. \ref{fig:general}, we illustrate how to select the hyperparameters of the variational circuit for a general function containing multiple zeros and singular points.

\subsection{Training Process}\label{ap:var}
%\subfile{sections/appendix/variational}
We proceed to illustrate the training method for the variational circuit proposed in this article. This process consists of iterative steps in which a loss function that measures how far the outcome is from the desired state is recursively minimized to obtain the optimal parameters.

Consider a function $f$ and $n$ qubits. The desired quantum state is
\begin{equation}
  \ket{\Psi(f)}_n = \sum_{l=0}^{2^n-1}f_l\ket{l},
\end{equation}
where the $f_l$ terms are discrete approximations to the objective function, as presented in Def.~\ref{df:f}.
%\newpage
Now, given a set of angles $\bm{\theta}$, our Ansatz $U(\bm{\theta})$ returns the state
\begin{equation}
  \ket{\Psi\left(\bm{\theta}\right)}_n=U(\bm{\theta})\ket{0},
\end{equation}
where $\ket{0}\equiv\ketzero{n}$, and the components of the obtained state are products of sines and cosines.
\\

Let us now introduce the mean squared error loss function, defined as
\begin{align*}
  L(n, f, \bm{\theta}) &= \frac{1}{2^n}\sum_{l=0}^{2^n-1}\left(f_l-\Psi_l\left(\bm{\theta}\right)\right)^2
\end{align*}
\begin{align}
=\frac{1}{2^n}\sum_{l=0}^{2^n-1}\left(f_l^2+\Psi_l\left(\bm{\theta}\right)^2-2f_l\Psi_l\left(\bm{\theta}\right)\right),
\end{align}

where $\Psi_l\left(\bm{\theta}\right):=\braket{l}{\Psi({\bm{\theta}})}_n$.
\\

This process aims to find the optimal parameters $\bm{\theta}$ for which $\ket{\Psi\left(\bm{\theta}\right)}_n$ approximates $\ket{\Psi(f)}_n$, which is equivalent to minimizing the loss function. Here, we use the gradient descent method~\cite{gradient} to do so.
\\

Then, given any angle $\theta_{l}^{(k-1)}$, with $k\in\{1, \dots, n\}$ and $l\in~\{0, 2^{k-1}-1\}$, its value gets updated after each training step in the following way:
\begin{equation}
  \theta_{l}^{(k-1)} = \theta_{l}^{(k-1)} - \gamma\frac{\partial L(n, f, \bm{\theta})}{\partial \theta_{l}^{(k-1)}},
\end{equation}
where $\gamma$ is the learning rate. Let us now compute the expression of the derivative:
\begin{align}
  \frac{\partial L(n, f, \bm{\theta})}{\partial \theta_{j}^{(k-1)}}=\frac{1}{2^n}\sum_{l=0}^{2^n-1}2\left(\Psi_l({\bm{\theta}})-f_l\right)\frac{\partial\Psi_l(\bm{\theta})}{\partial \theta_{j}^{(k-1)}}.
\end{align}

\begin{figure}[t!]
  \centering
  \includegraphics[width=1.0\columnwidth]{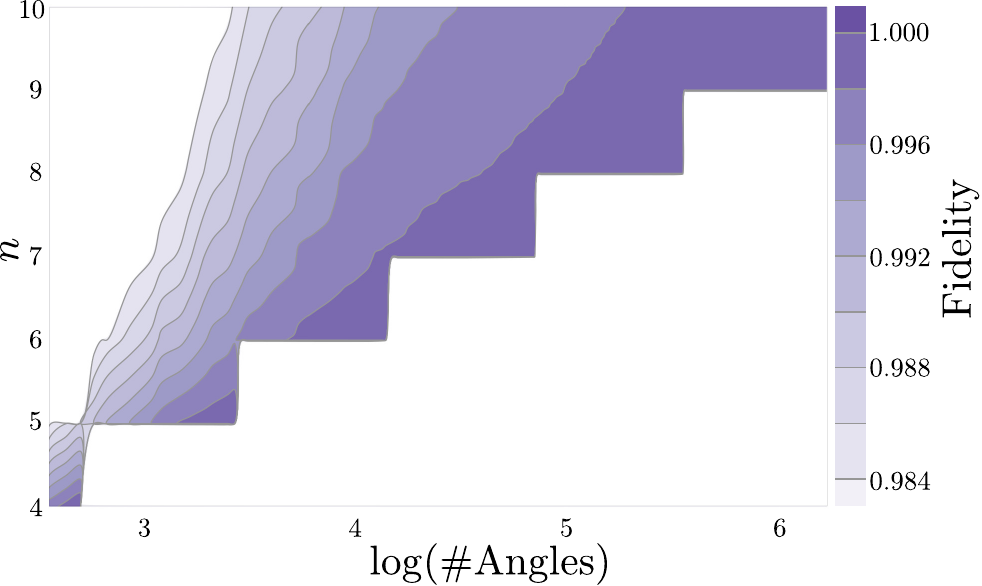}
  \caption{Fidelity of loading $f(x)=\sin{x}$, $x\in[0,3\pi/2]$, for different combinations of $n\in\{5, \dots, 10\}$, learning rate $\gamma=1.5$ and $p(k)$, with $k_0=3$. The axis corresponding to the total number of angles is in logarithmic scale.}
  \label{fig:sine-all}
\end{figure}

\begin{figure*}[t!]
\includegraphics[width=2\columnwidth]{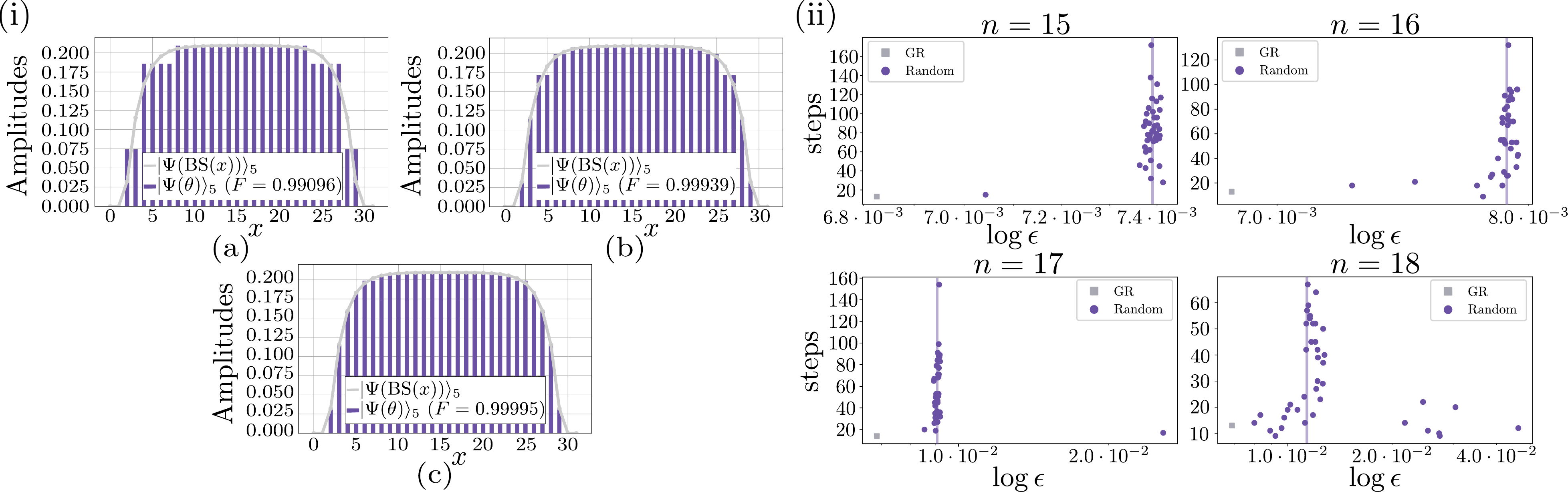}
\caption{(i) Simulations of $\ket{\Psi\left(\text{BS}(x)\right)}_5$ and $\ket{\Psi\left(\bm{\theta}\right)}_5$ for the Black-Scholes distribution, $n=5$, $k_0=2$, and different values of $p(k)$: (a) $p(k)=1$, (b) $p(k)=2$, and (c) $p(k)=3$. Parameters of the distribution: $K=45$ and $c=3$. Training parameters: $ k_0=2$, learning rate $\gamma=1.5$ and  tolerance of $10^{-9}$. We can appreciate that in all cases the circuit is able to capture the main features of the target state, consequently all fidelities are close to 1. (ii) Results of training the variational ans\"atze for a different number of qubits (n=15,16,17,18) to load the Black-Scholes distribution. We compare the training results (numbers of step and final infidelity) of 40 randomly initialized parameters versus the initialization with Grover-Rudolph method angles. These randomly initialized parameters are generated by drawing samples from a  uniform distribution between the values of 0 and $\pi$. As we can appreciate the initialization with Grover-Rudolph method always achieves the best performance, while most of random initializations get stuck around a fidelity value (vertical lines). Training parameters: $ k_0=2$, $p(k)=1$,  learning rate $\gamma=1.5$ and  tolerance of $10^{-9}$.}
\label{fig:bs_res}
\end{figure*}

Since $\Psi_l(\bm{\theta})$ is a product of sines and cosines, its partial derivative with respect to a given angle is
\begin{align}
  \frac{\partial\Psi_l(\bm{\theta})}{\partial \theta_{j}^{(k-1)}}=\begin{cases}
    -\frac{1}{2}\frac{\sin\left(\theta_{j}^{(k-1)}/{2}\right)}{\cos\left(\theta_{j}^{(k-1)}/{2}\right)}\Psi_l(\bm{\theta}) ~&\text{if}~j2^{n-k+1}\leq l< (j+1)2^{n-k},\\
    \frac{1}{2}\frac{\cos\left(\theta_{j}^{(k-1)}/{2}\right)}{\sin\left(\theta_{j}^{(k-1)}/{2}\right)}\Psi_l(\bm{\theta}) ~&\text{if}~(j+1)2^{n-k}\leq l< (j+1)2^{n-k+1},\\~
    0 ~&\text{otherwise}.
\end{cases}
\end{align}

Finally, different stopping criteria exist to put to an end the training process. A simple example is to fix a number of training steps. Another criteria, which is the one we consider for our numerical simulations, is to set a tolerance for the loss function. Therefore, when the difference between the cost function of two consecutive steps is less than the given tolerance, the process is considered satisfactory.

As a summary, in Algorithm {\ref{alg}} we have depicted the pseudocode corresponding to the training process.

\begin{algorithm}[h!]
\caption[singlelinecheck=false]{Algorithm for the training process.}\label{alg}
\begin{algorithmic}[1]
\Require $n$ qubits, normalized function $f$, objective state $\ket{\Psi(f)}_n$, circuit $U(\bm{\theta})$ with hyperparameters, stopping criteria and learning rate $\gamma$.
\State Initialize parameters $\bm{\theta}$
\While{stopping criteria is not met}
    \State $\ket{\Psi\left(\bm{\theta}\right)}_n \gets U(\bm{\theta})\ket{0}$
    \State Compute loss function $L(n, f, \bm{\theta})$
    \State $\bm{\theta} \gets \bm{\theta} - \gamma\frac{\partial L(n, f, \bm{\theta})}{\partial \bm{\theta}}$
\EndWhile
\end{algorithmic}
\end{algorithm}

%\begin{figure}[h]
%\includegraphics[width=1\columnwidth]{algorithm.png}
%\end{figure}
\subsection{Examples}
Let us now illustrate the behavior the variational circuit different density functions.

\subsubsection{Sine Function} As a first example, we have tested the variational circuit with the normalized sine function $\sin{x}$ in the domain $[0, \frac{3}{2}\pi]$. This example contains zeros in $x=0$ and $x=\pi$ and clearly does not satisfy the conditions in Th.~\ref{th:k0} or a possible generalization, since it is not even positive. In order to train the parametrized quantum circuit, we use the mean squared error loss function and the gradient descent as optimization method, with a learning rate $\gamma=1.5$. In Fig.~\ref{fig:results}\hyperref[fig:results]  (ii), we illustrate the arising loaded states of our trained circuits for different values of $p(k)=1$, 2, 3 and $k$ (the last case means the introduction of $k$ hyperparameters per zero of the function) and a system comprising $n=5$ qubits, with $k_0=2$. For the studied cases, our trained Ansatz is able to load the sine function with a fidelity larger than $0.97$. Additionally, in Fig.~\ref{fig:sine-all}, we show the resulting fidelity of the different combinations of the number of qubits and $p(k)$, for $n\in\{5, \dots, 10\}$ and $k_0=3$. 
\begin{table}[b!]
\caption{Numerical data for the simulations of $\ket{\Psi\left(\text{BS}(x)\right)}_{12}$ and $\ket{\Psi\left(\bm{\theta}\right)}_{12}$, $n=12$, and different values of $p(k)$, with $k_0=2$. The value \#Angles is the number of independent angles necessary to generate $\ket{\Psi\left(\bm{\theta}\right)}_{12}$, and \%Angles is in comparison with the full circuit. We are able to obtain large fidelities with only a few percentage of the original angles, which implies that with only a small portion of the initial parameter space, the circuit can approximately simulate the target state efficiently.}
\resizebox{1.\columnwidth}{!}{\begin{ruledtabular}
\begin{tabular}{cccc}
 $p(k)$&Fidelity&\#Angles&\%Angles\\
\hline
1 & 0.99303  & 33 & 0.806  \\
2 & 0.99838  & 52 & 1.269  \\
3 & 0.99890 & 70 & 1.709  \\
$k$ & 0.99913 & 142 & 3.468 \\
%$2k$ & 0.99893 & 262 & 6.398 \\
\end{tabular}
\end{ruledtabular}}
\label{tb:bs_res}
\end{table}

%%%%%%%%%%%%%%%%%%%%%%%%%%%%%%%%%%%%%%%%%%%%%

\subsubsection{Black–Scholes Distribution}\label{ap:bs}
As a second example, let us now apply the proposed variational circuit to the Black-Scholes distribution, which is given by
\begin{align}
\text{BS}(x)=\begin{cases}
    K-e^{-x}/s ~&\text{if}~-\log{(K s)}\leq x< 0,\\
    K-e^{x}/s ~&\text{if}~0<x\leq \log{(K s)},
    \label{eq:bs}
\end{cases}
\end{align}
with $s=Kc$, where $K$ and $c$ are the parameters. The zeros are found in $x=\pm\log{(K s)}$.
\\

\begin{table*}
\caption{Numerical results of training the variational ans\"atze for a different number of qubits (n=15,16,17,18) to load the Black-Scholes distribution. We compare the training results (numbers of step and final infidelity) of 40 randomly initialized parameters versus the initialization with Grover-Rudolph method angles. These randomly initialized parameters are generated by drawing samples from a uniform distribution between the values of 0 and $\pi$. Training parameters: $ k_0=2$, $p(k)=1$, learning rate $\gamma=1.5$ and  tolerance of $10^{-9}$.}
\resizebox{2.\columnwidth}{!}{\begin{ruledtabular}
\begin{tabular}{ccccccc}
 $n$&Fidelity (GR)&Steps (GR)& Avg. Fidelity (Random)& Avg. Steps (Random) & Max Fidelity (Random)& Dif. Fid. GR vs Avg. Random \\
\hline
15 & 0.99317  & 13 & 0.99261 & 82 & 0.99295 & 0.00056\\
16 & 0.99316  & 13 & 0.99211 & 59 &0.99271& 0.00105\\
17 & 0.99314 & 14 & 0.99052  &57 &0.99145& 0.00262\\
18 & 0.99309 & 13 & 0.98566 &31 &0.99199& 0.00743\\

\end{tabular}
\end{ruledtabular}}
\label{tb:bs_random}
\end{table*}
%First, in Fig.~\ref{fig:bs_disc}, we see the resulting circuit for the block of the uniformly-controlled rotation composed by 4 qubits and $p(k)=1$. As described previously, only the multi-controlled operations corresponding to the intervals containing zeros remain, the rest is clustered.

%\begin{figure}[H]
%\centering
%\includegraphics[width=\columnwidth]{bs_disc2.eps}
%\includegraphics[width=.6\columnwidth]{bs_disc2.png}
%\caption{Variational circuit corresponding to the 4-qubit block and $p(k)=1$ for the Black-Scholes distribution given by Eq.(\ref{eq:bs}), with $K=45$ and $c=3$. The function has two zeros of type 1 ($z1$) located in $x=\pm\log{(K s)}$, consequently we can chose our hyperparameter $k_0$ to be larger or equal than 2. Therefore, as illustrated, the clustered blocks will present two multi-controlled rotations corresponding to each zero and one single qubit rotation with the global clusterization angle. }
%\label{fig:bs_disc}
%\end{figure}

We benchmark the performance of our Ansatz for loading the Black-Scholes distribution, Eq.(\ref{eq:bs}), into a $5$-qubit system. We choose the parameters of the distribution $K=45$ and $c=3$, and for the training parameters, we consider, $k_0=2$, a learning rate of $\gamma=1.5$ and a tolerance of $10^{-9}$. In Fig.~\ref{fig:bs_res} (i), we depict the results of the from the numerical simulations of the discretized state $\ket{\Psi\left(\text{BS}(x)\right)}_5$ and the trained Ansatz $\ket{\Psi\left(\bm{\theta}\right)}_5$, for different values of $p(k)$ and $n=5$ qubits. We can appreciate that in all cases the circuit is able to capture the main features of the target state, consequently all fidelities are close to 1. We highlight that for $p(k)=3$ the resulting state and the objective one are almost equal, $\epsilon=5\mathrm{e}{-5}$.
\\

We also provide an analysis for larger sytems. In Table~\ref{tb:bs_res}, we depict the numerical results for $\ket{\Psi\left(\text{BS}(x)\right)}_{12}$ and $\ket{\Psi\left(\bm{\theta}\right)}_{12}$ in a system of 12-qubit system, with $k_0=2$, and different values of $p(k)$. We are able to obtain large fidelities with only a few percentage of the original angles, which implies that with only a small portion of the initial parameter space, the circuit can approximately simulate the target state efficiently.
%\newpage

%\newpage
Finally, in Fig~\ref{fig:bs_res} (ii) and Table~\ref{tb:bs_random} we analyze the performance of initializing the value of the parameters with the angles provided by the Grover-Rudolph method for different sizes of the system. We present the training results of the Ansatz initialized with several random angles sets and with the angles provided by the Grover-Rudolph method. We analyze the performance in terms of infidelity and number of steps of the training. We can observe how this initialization reduces drastically the number of steps, as well as enables us to achieve the largest fidelities.

%\begin{figure}[b!]
%\includegraphics[width=1.05\columnwidth]{scatter_bs.eps}
%\includegraphics[width=.9\columnwidth]{scatter_bs.png}
%\caption{Results of training the variational ans\"atze for a different number of qubits (n=15,16,17,18) to load the Black-Scholes distribution. We compare the training results (numbers of step and final infidelity) of 40 randomly initialized parameters versus the initialization with Grover-Rudolph method angles. These randomly initialized parameters are generated by drawing samples from a  uniform distribution between the values of 0 and $\pi$. As we can appreciate the initialization with Grover-Rudolph method always achieves the best performance, while most of random initializations get stuck around a fidelity value (vertical lines). Training parameters: $ k_0=2$, $p(k)=1$,  learning rate $\gamma=1.5$ and  tolerance of $10^{-9}$.}
%\label{fig:bs_random}
%\end{figure}

\section{Conclusions} In this article, we have considered the problem of loading real valued functions into a quantum computer, which is a major bottleneck for solving partial derivatives equations \cite{JAVI,CHILDS2,Zanger,RIPOLL}, computing Monte-Carlo integrations \cite{WOERNER,REBEN,MONTANARO} and quantum field theory \cite{KLCO,PRESKILL} and quantum machine learning~\cite{MARIA1,MARIA2,ML,ML2, WIEBE}. Firstly, inspired by the Grover-Rudolph algorithm \cite{GROVER}  without ancillas, we have analytically proven that the complexity for implementing our method on a $n$-qubit system scales as $\mathcal{O}(2^{k_0(\epsilon)})$, with $\epsilon$ the infidelity with respect to the exact state and $k_0(\epsilon)$ asymptotically independent of $n$. This reduction of two-qubit gates leads to a significant speedup, which allows us to implement quantum protocols involving data embeddings in large qubit systems. Additionally, we have generalized this method for functions containing a certain type of singularities, obtaining promising results for density functions with singularities that satisfy the expanded theorem conditions. Furthermore, we have proposed a variational Ansatz inspired in our previous protocol. We have observed that it can efficiently and accurately load functions with zeros and singularities. Our proposed Ansatz is tailored to the landscape of the function, providing an intuitive correlation between hyperparameters and expressability. Moreover,  our previous protocol allows us to define a suitable initial training angle set, which considerably improves the training process, avoiding barren plateaus and local minima. As a future work, tensor networks could be used to prove the quasi-optimality in the minimal number of hyperparameters in the variational Ansatz.
\\
%%%%%%%%%%%%%%%%%%%%%%%%%%%%%%%%%%%%%%%%%%%%%
\begin{acknowledgments}
We thank J. J. García-Ripoll for the useful discussions regarding the quasi optimality of
the variational Ansatz, T. Nguyen for the useful discussions regarding the Grover
Rudolph algorithm, and M. Garcia-de-Andoin for the discussion on the noise impact. The
authors acknowledge financial support from OpenSuperQ+100 (Grant No. 101113946)
of the EU Flagship on Quantum Technologies, as well as from the EU FET-Open project
EPIQUS (Grant No. 899368), also from Project Grant No. PID2021-125823NA-I00 595 and
Spanish Ramón y Cajal Grant No. RYC-2020-030503-I funded by
MCIN/AEI/10.13039/501100011033 and by “ERDF A way of making Europe” and “ERDF
Invest in your Future,” Basque Government through Grant No. IT1470-22 and the IKUR
Strategy under the collaboration agreement between Ikerbasque Foundation and BCAM
on behalf of the Department of Education of the Basque Government, as well as from
and UPV/EHU Ph.D. Grant No. PIF20/276.
\end{acknowledgments}
%%%%%%%%%%%%%%%%%%%%%%%%%%%%%%%%%%%%%%%%%%%%%
%%%%%%%%%%%%%%%%%%%%%%%%%%%%%%%%%%%%%%%%%%%%%

\section*{Appendix A: Gates Decomposition and Complexity}
\subsection*{A.1 Uniformly-Controlled Rotation} \label{ap:ucr}
\begin{figure}[h!]
  \centering
  \includegraphics[width=1\columnwidth]{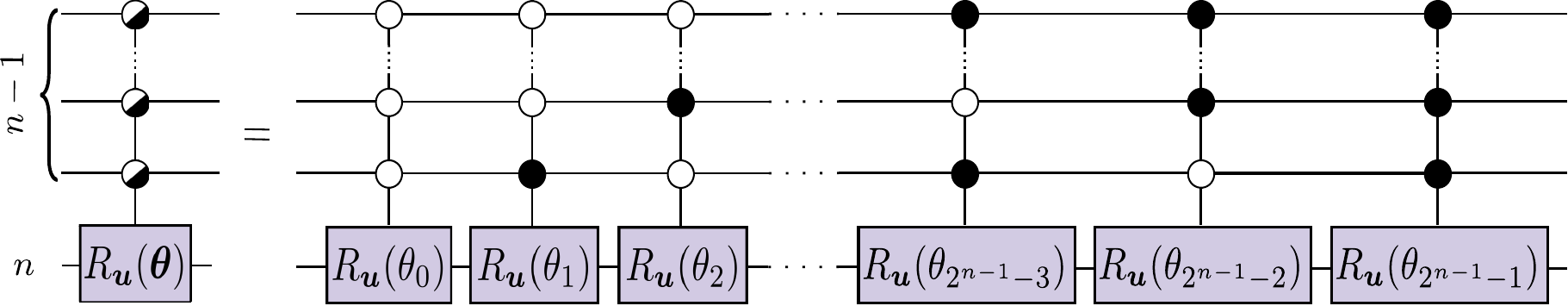}
  \caption{Circuit corresponding to the uniformly controlled rotation gate $F_n^{n-1}(\bm{u}, \bm{\theta})$, with $n-1$ control qubits, a target at the $n$-th qubit, an axis of rotation $\bm{u}$, and a set of angles $\bm{\theta}$.}
  \label{fig:ucrc}
\end{figure}

%\subfile{sections/appendix/ucr}
Let us consider $n-1$ control qubits, a target $n$-th qubit, an axis of rotation $\bm{u}$, and a vector of $2^{n-1}$ angles $\bm{\theta}$. Then, a uniformly controlled rotation $F_n^{n-1}(\bm{u}, \bm{\theta})$, depicted in Fig.~\ref{fig:ucrc}, is a sequence of multi-controlled rotations $\wedge_{n-1}(R_{\bm{u}}(\theta_i))$ comprising the $2^{n-1}$ combinations of control bits.

Analytically this gate can be expressed as

\begin{align}
  F_{n}^{n-1}(\bm{u}, \bm{\theta}) &=\prod_{i = 0}^{2^{n-1}-1}\left(\ket{i}\bra{i}\otimes R_{\bm{u}}(\theta_{i})+\left(\mathbbm{1}_{n-1\times n-1} - \ket{i}\bra{i}\right)\otimes\mathbbm{1}\right),
\end{align}
and since $\sum_{i = 0}^{2^n-1}\ket{i}\bra{i} = \mathbbm{1}_{n\times n}$ and $\braket{i}{j}=\delta_{ij}$, we can simplify it to

\begin{align}
  F_{n}^{n-1}(\bm{u}, \bm{\theta}) = \sum_{i = 0}^{2^{n-1}-1}\ket{i}\bra{i}\otimes R_{\bm{u}}(\theta_{i}).
\end{align}

%\newpage
%%%%%%%%%%%%%%%%%%%%%%%%%%%%%%%%%%%%%%%%%%%%%
\subsection*{A.2 Cost of Multi-Controlled Rotations}\label{ap:cost}

Let us now analyze the cost of implementing a multi-controlled rotation in terms of single and two-qubit gates.
%\subfile{sections/appendix/cost}
\begin{th1} \label{th:decomposition}
Let us consider an $n$-qubit system. Then, the multi-controlled rotation $\wedge_{n-1}\left(R_y(\theta)\right)$ can be decomposed employing two one-qubit controlled rotations $\wedge_{1}(R_y(\pm\theta/2))$ and two $\wedge_{n-2}(X)$, as illustrated in Fig.~\ref{fig:multi-dec}.

\end{th1}
\begin{proof}
First, let us see that $R_y(\theta) = XR_y(-\theta/2)XR_y(\theta/2)$:
\small
\begin{align}
    &XR_y(-\theta/2)X = \begin{pmatrix}0 & 1\\ 1 & 0\end{pmatrix}\cdot\begin{pmatrix} \cos(\nicefrac{\theta}{4}) & \sin(\nicefrac{\theta}{4})\\ -\sin\left(\nicefrac{\theta}{4}\right) & \cos(\nicefrac{\theta}{4})\end{pmatrix}\cdot\begin{pmatrix}0 & 1\\ 1 & 0\end{pmatrix} \nonumber\\& = \begin{pmatrix}0 & 1\\ 1 & 0\end{pmatrix}\cdot\begin{pmatrix} \sin(\nicefrac{\theta}{4}) & \cos(\nicefrac{\theta}{4})\\ \cos\left(\nicefrac{\theta}{4}\right) & -\sin(\nicefrac{\theta}{4})\end{pmatrix} = \begin{pmatrix} \cos(\nicefrac{\theta}{4}) & \sin(\nicefrac{\theta}{4})\\ -\sin\left(\nicefrac{\theta}{4}\right) & \cos(\nicefrac{\theta}{4})\end{pmatrix} = R_y(\theta/2)
\end{align}
\normalsize
Then,
\begin{equation}
  \label{eq:ry}
  XR_y(-\theta/2)XR_y(\theta/2) = R_y(\theta/2)R_y(\theta/2) = R_y(\theta).
\end{equation}

The intuition behind this decomposition is that when all the control states are in $\ket{1}$, we obtain the desired rotation, as illustrated in  Eq.~(\ref{eq:ry}). Also, when the controlled rotations are activated but the $\wedge_{n-2}(X)$ is not, we obtain the identity, since $R_y(-\theta/2)R_y(\theta/2)=\iden{1}$, as expected. The same occurs when the $\wedge_{n-2}(X)$ gates are activated but the controlled rotations are not. Finally, when none of the controls are triggered, the circuits in both sides are equal to $\iden{n}$.

%\vspace{0.1cm}
\noindent Now, let us check that the circuits are equivalent. On the left hand side, we have:
\begin{align}
    \wedge_{n-1}\left(R_y(\theta)\right) &= \ketone{(n-1)}\braone{(n-1)}\otimes R_y(\theta) \nonumber\\&+ \left(\iden{(n-1)}-\ketone{(n-1)}\braone{(n-1)}\right)\otimes\iden{1}.
\end{align}

\begin{figure}[b!]
    \centering
    \begin{tikzpicture}\node[scale=0.8]{
      \tikzset{operator/.append style={fill=violet!70!blue!20!} }
      \begin{quantikz}
    \lstick[wires=5]{$n-1$}
       \qw & \ctrl{1} & \qw \\
       \qw & \ctrl{1} & \qw \\
        & \ \vdots\  & \\
       \qw & \ctrl{-1} \vqw{1} & \qw \\
       \qw & \ctrl{1} & \qw \\
       \qw & \gate{R_y(\theta)} & \qw
     \end{quantikz} =
     \begin{quantikz}
       \qw & \qw &\ctrl{1} & \qw & \qw &\ctrl{1} & \qw\\
       \qw & \qw &\ctrl{1} & \qw & \qw &\ctrl{1} & \qw\\
        & \ \vdots\ &\ \vdots\  & & \ \vdots\ &\ \vdots\  & \\
       \qw & \qw & \ctrl{-1} \vqw{2} & \qw & \qw & \ctrl{-1} \vqw{2} & \qw\\
       \qw &\ctrl{1}& \qw & \qw &\ctrl{1}& \qw & \qw\\
       \qw & \gate{R_y(\theta/2)}& \targ{}& \qw & \gate{R_y(-\theta/2)}& \targ{} & \qw
        \end{quantikz}
       };
  \end{tikzpicture}
    \caption{Circuit showing the decomposition of $\wedge_{n-1}\left(R_y(\theta)\right)$ employing two controlled rotations $\wedge_{1}(R_y(\pm\theta/2))$ and two $\wedge_{n-2}(X)$ gates.}
    \label{fig:multi-dec}
\end{figure}
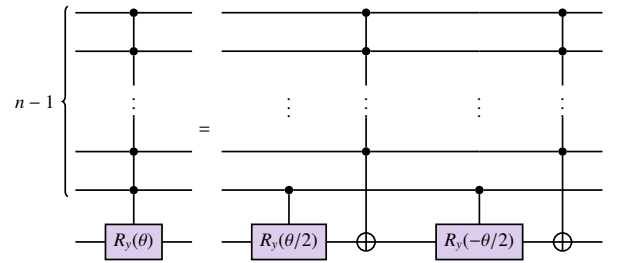
On the right-hand side:
%\small
\begin{align}
\small
    &\wedge_{n-2}(X)\wedge_{1}(R_y(-\theta/2))\wedge_{n-2}(X)\wedge_{1}(R_y(\theta/2)) \nonumber\\&\quad=\bigg[\ketone{(n-2)}\braone{(n-2)}\otimes\iden{1}\otimes X \\&\quad+ \left(\iden{(n-1)}-\ketone{(n-2)}\braone{(n-2)}\otimes\iden{1}\right)\otimes\iden{1}\bigg]\nonumber\\&\quad\cdot\bigg[\iden{(n-2)}\otimes\left(\ket{1}\bra{1}\otimes R_y(-\theta/2)+\ket{0}\bra{0}\otimes\iden{1}\right)\bigg]\nonumber\\
    &\quad\cdot\bigg[\ketone{(n-2)}\braone{(n-2)}\otimes\iden{1}\otimes X\\&\quad + \left(\iden{(n-1)}-\ketone{(n-2)}\braone{(n-2)}\otimes\iden{1}\right)\otimes\iden{1}\bigg]\nonumber\\&\quad\cdot\bigg[\iden{(n-2)}\otimes\left(\ket{1}\bra{1}\otimes R_y(\theta/2)+\ket{0}\bra{0}\otimes\iden{1}\right)\bigg],
\end{align}
\normalsize
where we have ignored the identities on the left hand side of the equation for the sake of simplicity.
If we expand the previous expression, we obtain the following:
%\footnotesize
\begin{align}
    &\bigg[\ket{1\cdots1}\bra{1\cdots1}\otimes XR_y(-\theta/2) \\&\quad+ \ket{1\cdots10}\bra{1\cdots10}\otimes X + F(-\theta/2)\bigg]\nonumber\\
    &\cdot \bigg[\ket{1\cdots1}\bra{1\cdots1}\otimes XR_y(\theta/2)\\&\quad + \ket{1\cdots10}\bra{1\cdots10}\otimes X + F(\theta/2)\bigg],
\end{align}
\normalsize
where we have defined the projector
\begin{align}
    F(\alpha) &\equiv \left[\left(\iden{(n-1)}-\ket{1\cdots1}\bra{1\cdots1}-\ket{1\cdots10}\bra{1\cdots10}\right)\otimes\iden{1}\right]\nonumber\\
    &\cdot\left[\iden{(n-2)}\otimes\left(\ket{1}\bra{1}\otimes R_y(\alpha)+\ket{0}\bra{0}\otimes\iden{1}\right)\right] \nonumber\\
    & = \iden{(n-2)}\otimes\ket{1}\bra{1}\otimes R_y(\alpha) + \iden{(n-2)}\otimes\ket{0}\bra{0}\otimes \iden{1} \nonumber\\
    &- \ket{1\cdots10}\bra{1\cdots10}\otimes\iden{1} -  \ket{1\cdots1}\bra{1\cdots1}\otimes R_y(\alpha).
\end{align}
It is straightforward to see that both terms \small$\ket{1\cdots1}\bra{1\cdots1}\otimes XR_y(\alpha)\cdot F(\alpha)$\normalsize\ and \small$\ket{1\cdots10}\bra{1\cdots10}\otimes X\cdot F(\alpha)$\normalsize\ are equal to $0$, due to the projector $F(\alpha)$. Then, we have
%\small
\begin{align}
    &\wedge_{n-2}(X)\wedge_{1}(R_y(-\theta/2))\wedge_{n-2}(X)\wedge_{1}(R_y(\theta/2)) \nonumber\\&\quad= \ket{1\cdots1}\bra{1\cdots1}\otimes R_y(\theta)+\ket{1\cdots10}\bra{1\cdots10}\otimes \iden{1}\\&\quad +F(-\theta/2)F(\theta/2),
\end{align}
\normalsize
where we have used that {\small $XR_y(-\theta/2)XR_y(\theta/2) = R_y(\theta)$} and {\small $XX=\iden{1}$}. Let us now compute the last term:
%\small
\begin{align}
    &F(-\theta/2)F(\theta/2) \nonumber\\ &\quad= \iden{(n-2)}\otimes\ket{1}\bra{1}\otimes R_y(-\theta/2)R_y(\theta/2)\\ &\quad - \ket{1\cdots1}\bra{1\cdots1}\otimes R_y(-\theta/2)R_y(\theta/2) \nonumber\\ &\quad +\iden{(n-2)}\otimes\ket{1}\bra{1}\otimes\iden{1} - \ket{1\cdots10}\bra{1\cdots10}\otimes \iden{1}\\ &\quad - \ket{1\cdots10}\bra{1\cdots10}\otimes \iden{1} \nonumber +\ket{1\cdots10}\bra{1\cdots10}\otimes \iden{1}\\ &\quad - \ket{1\cdots1}\bra{1\cdots1}\otimes R_y(-\theta/2)R_y(\theta/2)\nonumber\\ &\quad +\ket{1\cdots1}\bra{1\cdots1}\otimes R_y(-\theta/2)R_y(\theta/2)\nonumber\\ &\quad = \left(\iden{(n-1)}-\ket{1\cdots1}\bra{1\cdots1}-\ket{1\cdots10}\bra{1\cdots10}\right)\otimes\iden{1},
\end{align}
\normalsize
where we have used that $R_y(-\theta/2)R_y(\theta/2)=\iden{1}$ and
\begin{equation}
  \iden{(n-2)}\otimes\ket{1}\bra{1}\otimes\iden{1} + \iden{(n-2)}\otimes\ket{0}\bra{0}\otimes\iden{1}=\iden{(n-1)}\otimes\iden{1}.
\end{equation}
Finally,
\begin{align}
    &\wedge_{n-2}(X)\wedge_{1}(R_y(-\theta/2))\wedge_{n-2}(X)\wedge_{1}(R_y(\theta/2)) \nonumber\\&\quad = \ket{1\cdots1}\bra{1\cdots1}\otimes R_y(\theta)+\ket{1\cdots10}\bra{1\cdots10}\otimes \iden{1} \nonumber\\&\quad +\left(\iden{(n-1)}-\ket{1\cdots1}\bra{1\cdots1}-\ket{1\cdots10}\bra{1\cdots10}\right)\otimes\iden{1}\nonumber\\&\quad = \ket{1\cdots1}\bra{1\cdots1}\otimes R_y(\theta) + \left(\iden{(n-1)}-\ket{1\cdots1}\bra{1\cdots1}\right)\otimes\iden{1}\nonumber\\&\quad= \wedge_{n-1}\left(R_y(\theta)\right),
\end{align}
as we wanted to prove.
\end{proof}

\begin{cor1}
 \label{cor:mqg}
The order of complexity of implementing the $\wedge_{n-1}\left(R_y(\theta)\right)$ gate is linear in $n$, for $n>6$. In fact, $\wedge_{n-1}\left(R_y(\theta)\right)$ can be decomposed with $80n-398$ two-qubit gates.
\end{cor1}
\begin{proof}
Following Th. \ref{th:decomposition}, we know that the multi-controlled rotation $\wedge_{n-1}\left(R_y(\theta)\right)$ can be decomposed into two one-qubit controlled rotations,  $\wedge_{1}\left(R_y(\theta/2)\right)$ and $\wedge_{1}\left(R_y(-\theta/2)\right)$, and two $\wedge_{n-2}\left(X\right)$. Therefore, we need to study the number of two-qubit gates necessary to implement the last two gates $\wedge_{n-2}\left(X\right)$.

Corollary $7.4$ from \cite{BARENCO} states that, for $n>6$, $\wedge_{n-2}\left(X\right)$ can be realized with $8(n-5)\wedge_{2}\left(X\right)$. Then, we reduce the problem  to obtain the cost of the Toffoli gate, $\wedge_{2}\left(X\right)$.

Lemma 6.1 from the previous reference \cite{BARENCO} states that any one-qubit unitary, $\wedge_{2}\left(U\right)$ can be implemented with 5 two-qubit gates comprising two CNOTs, two $\wedge_{1}\left(V\right)$, and one $\wedge_{1}\left(V^\dagger\right)$, with $V^2=U$. In our case, we have $U=X$. Then, we can choose $V = \frac{1-i}{2}(\iden{1}+iX)$. We provide the exact decomposition of the Toffoli gate using the previous description in Fig.~\ref{fig:2x-dec}.

\begin{figure}[t!]
    \centering
    \begin{tikzpicture}\node[scale=1.]{
      \tikzset{operator/.append style={fill=violet!70!blue!20!} }
      \begin{quantikz}
       \qw & \ctrl{1} & \qw \\
       \qw & \ctrl{1} & \qw \\
       \ghost{R_y(-\frac{\pi}{4})}\qw & \targ{} & \qw
     \end{quantikz} =
     \begin{quantikz}
     \qw & \qw  & \ctrl{1} & \qw & \ctrl{1}& \ctrl{2}& \qw\\
     \qw & \ctrl{1}  &\targ{}&  \ctrl{1} & \targ{}& \qw & \qw\\
     \qw & \gate{V}& \qw&\gate{V^\dagger} & \qw & \gate{V}& \qw
        \end{quantikz}
       };
  \end{tikzpicture}
    \caption{Circuit of the decomposition of the Toffoli gate, $\wedge_{n-2}\left(X\right)$ using five two-qubit controlled operations, where $V^2=X$.}
    \label{fig:2x-dec}
\end{figure}
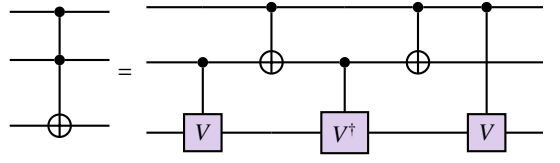

Finally, if $n>6$, then the number of gates required for decomposing $\wedge_{n-1}\left(R_y(\theta)\right)$, considering only two-qubit operators, is
\begin{align}
    \#{\text {gates}}\left(\wedge_{n-1}\left(R_y(\theta)\right)\right) = 2 + 2\cdot\left(8(n-5)\cdot5\right) = 80n-398.
\end{align}
\end{proof}
%%%%%%%%%%%%%%%%%%%%%%%%%%%%%%%%%%%%%%%%%%%%%

\section*{Appendix B: Theorem Proofs}\label{ap:proofs}
\subsection*{B.1 Relation Between the Difference in the Angles and the Target Function}\label{ap:angles}
%\subfile{sections/appendix/angles}
\begin{th1}
\label{th:2derlog}
Let $f$ be a continuous function such that $f:[0, 1]\to\mathbb{R}^+$ and consider a block comprising a uniformly controlled rotation of $k$ qubits. Then, the difference between two contiguous angles is bounded by the second derivative of the logarithm of $f$ in the following way:
\begin{equation}
  \left|\theta^{(k-1)}_{l+1}-\theta^{(k-1)}_l\right| \leq \frac{\delta_k^2}{4}\max_{y'\in[l\delta_k, (l+1)\delta_k]}\left|\left.\left(\partial^2_{y}\log f(y)\right)\right\rvert_{y=y'}\right|,
\end{equation}
where $\delta_k=\frac{1}{2^{k-1}}$, $l\in\{1, \dots, 2^{k-1}-1\}$.
\end{th1}
\begin{proof}
The discrete expression of the angles for a block comprised of uniformly controlled rotation with $k$ qubits, first introduced in~\cite{GROVER}, is given by
\begin{equation}
    \theta^{(k-1)}_l = 2\arccos\left(\sqrt{\frac{\int_{l\delta_k}^{(l+1/2)\delta_k}f(x)dx}{\int_{l\delta_k}^{(l+1)\delta_k}f(x)dx}}\right),
\end{equation}
with $\delta_k=\frac{1}{2^{k-1}}$ and $l\in\{1, \dots, 2^{k-1}-1\}$.
We can define the continuous extension of the previous function as
\begin{equation}
    \theta^{(k-1)} (y) \vcentcolon= 2\arccos\left(\sqrt{\frac{\int_{y}^{y +\delta_k/2}f(x)dx}{\int_{y}^{y +\delta_k}f(x)dx}}\right),
    \label{eq:thetas}
\end{equation}
where the original expression can be recovered replacing $y=l\delta_k$.

Now, given a function $g$ and a displacement $\mu$, we define the numerical derivative of $g$ with respect to $\mu$ as
\begin{equation}
    \npartial{y}{\mu}g(y) \vcentcolon= \frac{g(y+\mu)-g(y)}{\mu}.
    \label{eq:num-der}
\end{equation}
Notice that for $\mu \rightarrow 0$, $\npartial{y}{\mu}g(y) \rightarrow \partial_yg(y)$, the usual derivative. Then, for two consecutive angles, we have
\begin{equation}
    \left|\theta^{(k-1)}_{l+1}-\theta^{(k-1)}_l\right| = \left|\left. \delta_k\left( \npartial{y}{\delta_k}\theta^{(k-1)}(y)\right)\right\rvert_{y=l\delta_k}\right|.
\end{equation}

Now, we require a connection between the difference in the angles and the partial derivative of Eq.(\ref{eq:thetas}). The Mean Value Theorem guarantees that the numerical derivative, which corresponds to the slope of the straight line connecting $\theta^{(k-1)}_l$ and $\theta^{(k-1)}_{l+1}$, is constrained by the maximum absolute value of the exact derivative in that interval. Then, we have
\begin{align}\label{eq:der-ineq}
    \left|\theta^{(k-1)}_{l+1}-\theta^{(k-1)}_l\right| \leq \delta_k\max_{y'\in[l\delta_k, (l+1)\delta_k]}\left|\left.\left(\partial_{y}\theta^{(k-1)}(y)\right)\right\rvert_{y=y'}\right|.
\end{align}
Let us now develop the term of the exact derivative of the function $\theta^{(k-1)}(y)$:
\small
\begin{equation}
    \partial_{y}\theta^{(k-1)}(y) 
\end{equation}
\begin{equation*}
 = \frac{\left(f(y+\delta_k)-f(y)\right)\int_y^{y+\delta_k/2}f(x)dx - \left(f(y+\delta_k/2)-f(y)\right)\int_y^{y+\delta_k}f(x)dx}{\int_y^{y+\delta_k}f(x)dx \sqrt{\int_y^{y+\delta_k/2}f(x)dx \int_{y+\delta_k/2}^{y+\delta_k}f(x)dx}}.
\end{equation*}
\\

\normalsize
By introducing $\left(f(y+\delta_k/2)-f(y+\delta_k/2)\right)$ in the first term and reorganizing the expression, we have:
\small
\begin{align}
    &\partial_{y}\theta^{(k-1)}(y)\nonumber\\&\quad= \frac{\left(f(y+\delta_k)-f(y+\delta_k/2)\right)\int_y^{y+\delta_k/2}f(x)dx }{\int_y^{y+\delta_k}f(x)dx \sqrt{\int_y^{y+\delta_k/2}f(x)dx \int_{y+\delta_k/2}^{y+\delta_k}f(x)dx}} \nonumber\\&\quad -\frac{ \left(f(y+\delta_k/2)-f(y)\right)\left(\int_y^{y+\delta_k}f(x)dx - \int_y^{y+\delta_k/2}f(x)dx\right)}{\int_y^{y+\delta_k}f(x)dx \sqrt{\int_y^{y+\delta_k/2}f(x)dx \int_{y+\delta_k/2}^{y+\delta_k}f(x)dx}} \nonumber\\
    &\quad= \frac{\left(f(y+\delta_k)-f(y+\delta_k/2)\right)\int_y^{y+\delta_k/2}f(x)dx }{\int_y^{y+\delta_k}f(x)dx \sqrt{\int_y^{y+\delta_k/2}f(x)dx \int_{y+\delta_k/2}^{y+\delta_k}f(x)dx}}
    \nonumber\\
    &\quad - \frac{ \left(f(y+\delta_k/2)-f(y)\right)\int_{y+\delta_k/2}^{y+\delta_k}f(x)dx}{\int_y^{y+\delta_k}f(x)dx \sqrt{\int_y^{y+\delta_k/2}f(x)dx \int_{y+\delta_k/2}^{y+\delta_k}f(x)dx}}\nonumber\\
    &\quad= \frac{\sqrt{\int_y^{y+\delta_k/2}f(x)dx \int_{y+\delta_k/2}^{y+\delta_k}f(x)dx}}{\int_y^{y+\delta_k}f(x)dx}\nonumber\\ &\quad \cdot\left(\frac{f(y+\delta_k)-f(y+\delta_k/2)}{\int_{y+\delta_k/2}^{y+\delta_k}f(x)dx}-\frac{f(y+\delta_k/2)-f(y)}{\int_y^{y+\delta_k/2}f(x)dx}\right),
\end{align}
\normalsize
where the first term is positive. Now, since $\left(\sqrt{a}-\sqrt{b}\right)^2=a+b-2\sqrt{ab}\geq 0$, we have the well-known inequality for the geometric and arithmetic means $\sqrt{ab}\leq\frac{a+b}{2}$. Then, if we select $a=\frac{\int_y^{y+\delta_k/2}f(x)dx}{\int_y^{y+\delta_k}f(x)dx}$ and $b=\frac{\int_{y+\delta_k/2}^{y+\delta_k}f(x)dx}{\int_y^{y+\delta_k}f(x)dx}$, it leads to
\begin{equation*}
  \frac{\sqrt{\int_y^{y+\delta_k/2}f(x)dx \int_{y+\delta_k/2}^{y+\delta_k}f(x)dx}}{\int_y^{y+\delta_k}f(x)dx} 
\end{equation*}
\begin{equation}
    \leq \frac{1}{2}\frac{\int_y^{y+\delta_k/2}f(x)dx + \int_{y+\delta_k/2}^{y+\delta_k}f(x)dx}{\int_y^{y+\delta_k}f(x)dx}=\frac{1}{2}.
\end{equation}
%\newpage
With this result, we get the following inequality for the absolute value of the derivative of $\theta^{(k-1)}(y)$:
\begin{align}
    \label{eq:angle-ineq}
    \left|\partial_{y}\theta^{(k-1)}(y)\right| \leq \frac{1}{2}\left|\frac{f(y+\delta_k)-f(y+\delta_k/2)}{\int_{y+\delta_k/2}^{y+\delta_k}f(x)dx}-\frac{f(y+\delta_k/2)-f(y)}{\int_y^{y+\delta_k/2}f(x)dx}\right|.
\end{align}

By using again the numerical derivative defined in Eq.~(\ref{eq:num-der}), we can simplify the previous expression in the following way:
\small
\begin{align}
    \frac{f(y+\delta_k)-f(y+\delta_k/2)}{\int_{y+\delta_k/2}^{y+\delta_k}f(x)dx}-\underbrace{\frac{f(y+\delta_k/2)-f(y))}{\int_y^{y+\delta_k/2}f(x)dx}}_{\vcentcolon=h(y)} \\= h(y+\delta_k/2) - h(y) = \frac{\delta_k}{2}\npartial{y}{\delta_k/2}h(y).
\end{align}
\normalsize
Also, since $f$ must be integrable, its primitive $F$ exists. Then,
\begin{align}
    h(y)= \frac{f(y+\delta_k/2)-f(y)}{\int_y^{y+\delta_k/2}f(x)dx}=\frac{f(y+\delta_k/2)-f(y)}{F(y+\delta_k/2)-F(y)} = \frac{\npartial{y}{\delta_k/2}f(y)}{\npartial{y}{\delta_k/2}F(y)}.
\end{align}
Let us now plug it into the inequality
\begin{align*}
    \left|\partial_{y}\theta^{(k-1)}(y)\right| \leq \frac{1}{2}\left|\frac{\delta_k}{2}\cdot\npartial{y}{\delta_k/2}\left(\frac{\npartial{y}{\delta_k/2}f(y)}{\npartial{y}{\delta_k/2}F(y)}\right)\right|
\end{align*}
\begin{equation}
     = \frac{\delta_k}{4}\left|\npartial{y}{\delta_k/2}\left(\frac{\npartial{y}{\delta_k/2}f(y)}{\npartial{y}{\delta_k/2}F(y)}\right)\right|.
\end{equation}

If we follow the same argument as in Eq.~(\ref{eq:der-ineq}), we have that the term of the numerical derivative is upper bounded by the exact derivative, in the absolute value, which corresponds to the second derivative of the function's logarithm. Hence, the following inequality holds
\begin{equation}
  \left|\partial_{y}\theta^{(k-1)}(y)\right|_{[l\delta_k, (l+1)\delta_k]} \leq \frac{\delta_k}{4}\max_{y'\in[l\delta_k, (l+1)\delta_k]}\left|\left.\left(\partial^2_{y}\log f(y)\right)\right\rvert_{y=y'}\right|.
\end{equation}
Finally, by plugging this result in Eq.~(\ref{eq:der-ineq}), we obtain
\begin{align*}
    \left|\theta^{(k-1)}_{l+1}-\theta^{(k-1)}_l\right| \leq\max_{y'\in[l\delta_k, (l+1)\delta_k]}\left|\frac{\delta_k^2}{4}\max_{y'\in[l\delta_k, (l+1)\delta_k]}\left|\left.\left(\partial^2_{y}\log f(y)\right)\right\rvert_{y=y'}\right|\right|
\end{align*}
\begin{equation}
=\frac{\delta_k^2}{4}\max_{y'\in[l\delta_k, (l+1)\delta_k]}\left|\left.\left(\partial^2_{y}\log f(y)\right)\right\rvert_{y=y'}\right|,
\end{equation}

as we wanted to proof.
%Finally, if $\left|\partial_y^2\log f(y)\right|\leq \eta$, with $\eta\geq 0$, we obtain
%\begin{align}
%     \left|\partial_{y}\theta^{(k-1)}(y)\right| \leq \frac{\delta_k}{4}\eta\implies \left|\theta^{(n-1)}_{l+1}-\theta^{(k-1)}_l\right|\leq\frac{\delta_k^2}{4}\eta,
%\end{align}
\end{proof}
\begin{cor1}
  \label{cor:angle-bound}
  Let $f$ be a continuous function such that $f:[0, 1]\to\mathbb{R}^+$ and consider a block of uniformly controlled rotations of $k$ qubits. Then, if $\exists \eta \geq 0$ such that $\left|\partial^2_{y}\log f(y)\right| \leq \eta ~\forall y\in[0, 1]$,
  \begin{equation}
    \left|\theta^{(k-1)}_{l+1}-\theta^{(k-1)}_l\right|\leq \frac{\delta_k^2}{4}\eta.
  \end{equation}
\end{cor1}
\begin{cor1}
For a non-standardized function $f:[x_{\text{min}}, x_{\text{max}}]\to\mathbb{R}^+$, the result in Corollary \ref{cor:angle-bound} holds with the modification in the bound
\begin{equation}
    \left|\theta^{(k-1)}_{l+1}-\theta^{(k-1)}_l\right| \leq \frac{\delta_k^2\cdot\eta}{4L^2},
\end{equation}
where $L=x_{\text{max}}-x_{\text{min}}$ and $\delta_k=\frac{L}{2^{k-1}}$.
\end{cor1}
\begin{proof}
The change of variables that map the $x'\in[0, 1]$ with $x\in[x_{\text{max}}-x_{\text{min}}]$ is given by
\begin{equation}
    x = x_{\text{min}} + x'L.
\end{equation}
Now, by using the chain rule, we obtain:
\begin{equation}
    \frac{\partial f}{\partial y'} = \frac{\partial f}{\partial y}\frac{\partial y'}{\partial y'} =  \frac{\partial f}{\partial y}L \implies \left|\partial^2_{y}\log f(y)\right| \leq \frac{\tilde{\eta}}{L^2}.
\end{equation}
\end{proof}
\begin{cor1}\label{cor:rep}
  Let $\eta$ be such that $\left|\partial^2_{y}\log f(y)\right| \leq \eta ~\forall y\in[0, 1]$. Then, the difference between any two angles is
  \begin{equation}
    \left|\theta^{(k-1)}_{l}-\theta^{(k-1)}_{l'}\right|\leq\frac{\delta_k}{4}\eta,
  \end{equation}
$\forall l, l'\in\{1, \dots, 2^{k-1}\}$.
\end{cor1}
\begin{proof}
  The worst scenario is when $l=1$ and $l'=2^{k-1}$. In this case, by using the triangular inequality, we have
  \begin{equation}
    \left|\theta^{(k-1)}_{1}-\theta^{(k-1)}_{2^{k-1}}\right|\leq\left|\theta^{(k-1)}_{1}-\theta^{(k-1)}_{2}\right|+\cdots+\left|\theta^{(k-1)}_{2^{k-1}-1}-\theta^{(k-1)}_{2^{k-1}}\right|\leq\frac{\delta_k}{4}\eta,
  \end{equation}
  since there are less than $\delta_k^{-1}$ elements in the sum.
\end{proof}

In this situation, we can define the representative angle $\tilde{\theta}^{(k-1)}$ for the clustering process as the one corresponding to the middle part of the interval, satisfying
\begin{equation}\label{eq:representative_ap}
  \left|\tilde{\theta}^{(k-1)}-\theta^{(k-1)}_{l}\right|\leq\frac{\delta_k}{8}\eta \quad\forall l\in\{0, 1, \dots, 2^{k-1}-1\}.
\end{equation}
%%%%%%%%%%%%%%%%%%%%%%%%%%%%%%%%%%%%%%%%%%%%%
\subsection*{B.2 Error Bound of the Algorithm}
%\subfile{sections/appendix/fidelity}
\subsubsection{Relation Between the Difference in the Angles and the Fidelity}\label{ap:fid}
Let us consider a system with $n$ qubits, thus the unitary gate to prepare the quantum state representing the target density function can be written in terms of the blocks as
\begin{equation}
\mathfrak{U}_n = \mathcal{U}_{n-1}\left(\bm{\theta}^{(n-1)}\right)\cdots\mathcal{U}_{0}\left(\bm{\theta}^{(0)}\right),
\end{equation}
where we define
\begin{align}
\mathcal{U}_{k-1}\left(\bm{\theta}^{(k-1)}\right) &\coloneqq F_{k}^{k-1}(\bm{y}, \bm{\theta}^{(k-1)})\otimes\iden{(n-k)}.
\end{align}

Let $\tilde{\mathfrak{U}}_n$ denote the operation $\mathfrak{U}_n$ given a representative with an error $\eta_k$ between the angles for each block, i.e. $|\theta_l^{(k-1)} -\tilde{\theta}^{(k-1)}|\leq \eta_k$ for $l=0, \dots, 2^{k-1}-1$ and $k=1, \dots, n$.
\begin{th1}\label{th:fid}
Consider a system of $n$ qubits and an error $\eta_k$ between any angle of the $k$-th block and its representative such that $\eta_k\leq\pi$, with $k=1, \dots, n$. Then, the fidelity between the final states with and without clustering, $F = |\brazero{n}\mathfrak{U}_n^\dagger \tilde{\mathfrak{U}}_n\ketzero{n}|^2$, satisfies
  \begin{equation}
  F \geq \prod_{k=1}^n\cos^{2}\left(\eta_k/2\right).
  \end{equation}
\end{th1}

\begin{proof}\leavevmode
  We use induction to prove the inequality. Hence, we start with the elemental case of $n=1$ and later proceed assuming it is satisfied for $n-1$ and check if it holds for $n$:
  \begin{itemize}%[leftmargin=*]
      \item $n=1$:
      \begin{align}
          \bra{0}\mathfrak{U}_1^\dagger \tilde{\mathfrak{U}}_1\ket{0} =& \bra{0}R_y^\dagger(\theta^{(0)}) R_y(\tilde{\theta}^{(0)})\ket{0}\nonumber\\=& \bra{0}R_y^\dagger\left(\theta^{(0)} - \tilde{\theta}^{(0)}\right)\ket{0} = \cos\left(\frac{\theta^{(0)} - \tilde{\theta}^{(0)}}{2}\right)
      \end{align}
      Notice that the angles, given by %Eq. \ref{eqn:angles},
      \begin{align}
      %\label{eqn:angles}
        \theta^{(n-1)}_l = 2\arccos\left(\sqrt{\frac{\int_{x_\text{min}+ l\delta_n}^{x_\text{min} + (l+1/2)\delta_n}f(x)dx}{\int_{x_\text{min}+ l\delta_n}^{x_\text{min} + (l+1)\delta_n}f(x)dx}}\right),
\end{align}
      take values between $2\arccos(1)=0$ and $2\arccos(0)=\pi$. Then, since the cosine is a decreasing function in the interval $[0, \pi/2]$ and $\eta\leq\pi$, we have that
      \begin{align}
          |\theta^{(0)} - \tilde{\theta}^{(0)}|\leq \eta_1 &\implies
          \cos\left(\frac{\theta^{(0)} - \tilde{\theta}^{(0)}}{2}\right) \geq \cos\left(\eta_1/2\right) \nonumber\\&\implies F \geq \cos^{2}\left(\eta_1/2\right).
      \end{align}
      \item $n>1$: Assume the condition holds for $n-1$. We aim to recover the expression of $\brazero{n-1}\mathfrak{U}_{n-1}^\dagger \tilde{\mathfrak{U}}_{n-1}\ketzero{n-1}$ so we can use the induction hypothesis. Therefore, as we did for $n=1$, we sandwich the operator
      \begin{equation}
          \left(F_{n}^{n-1}(\bm{y}, \bm{\theta}^{(n-1)})\right)^\dagger F_{n}^{n-1}(\bm{y}, \bm{\tilde{\theta}}^{(n-1)})
      \end{equation}
      with the state $\ketzero{1}$ at each side. From now on during this proof, to simplify the notation,  we denote the gate $F_{k}^{k-1}(\bm{y}, \bm{\tilde{\theta}}^{(k-1)})$ as $U_{k-1}(\bm{\theta}^{(k-1)})$. Then, for $n$:
      %{\setlength{\mathindent}{0cm}
      \small
      \begin{align}
      %\allowdisplaybreaks[1]
      &\brazero{n}\mathfrak{U}_n^\dagger \tilde{\mathfrak{U}}_n\ketzero{n} \nonumber\\ & =\brazero{n} \mathcal{U}^\dagger_{0}\left(\theta^{(0)}\right)\cdots \mathcal{U}^\dagger_{n-1}\left(\bm{\theta}^{(n-1)}\right)\mathcal{U}_{n-1}\left(\bm{\tilde{\theta}}^{(n-1)}\right)\cdots
      \mathcal{U}_{0}\left(\tilde{\theta}^{(0)}\right)\ketzero{n}\nonumber\\
      &=\brazero{(n-1)}\otimes\bra{0} \left(U_0^\dagger(\theta^{(0)})\otimes\iden{(n-1)}\right)\cdots
      \left(U_{n-2}^\dagger(\bm{\theta}^{(n-2)})\otimes \mathbbm{1}\right)
      \cdot\left(U_{n-1}^\dagger(\bm{\theta}^{(n-1)})\right)\nonumber\\&\cdot\left(U_{n-1}(\bm{\tilde{\theta}}^{(n-1)})\right)
      \left(U_{n-2}(\bm{\tilde{\theta}}^{(n-2)})\otimes \mathbbm{1}\right)\cdots \left(U_0(\tilde{\theta}^{(0)})\otimes\iden{(n-1)}\right)\ketzero{(n-1)}\otimes\ket{0}\nonumber\\
      &=\brazero{(n-1)}\left(U_0^\dagger(\theta^{(0)})\otimes\iden{(n-2)}\right)\cdots \left(U_{n-2}^\dagger(\bm{\theta}^{(n-2)})\right)\nonumber\\&
      \cdot\left(\sum_{i_1, \ldots, i_{n-1} = 0}^1 \ket{i_1\ldots i_{n-1}}\bra{i_1\ldots i_{n-1}}\cdot \bra{0}  R_y^\dagger(\theta_{i_1\ldots i_{n-1}}^{(n-1)}-\tilde{\theta}^{(n-1)})\ket{0}\right)\nonumber\\
      &\cdot\left(U_{n-2}(\bm{\tilde{\theta}}^{(n-2)})\right)\cdots \left(U_0(\tilde{\theta}^{(0)})\otimes\iden{(n-2)}\right)\ketzero{(n-1)}
    \end{align}
    \normalsize
    Notice that we now we can substitute the rotation terms in the following way:
    \begin{align*}
      \bra{0}  R_y^\dagger(\theta_{i_1\ldots i_{n-1}}^{(n-1)}-\tilde{\theta}^{(n-1)})\ket{0} 
    \end{align*}
\begin{equation}
    = \cos\left(\frac{\theta_{i_1\ldots i_{n-1}}^{(n-1)}-\tilde{\theta}^{(n-1)}}{2}\right)\geq\cos\left({\eta_n/2}\right),
\end{equation}
    
    since $\theta_{i_1\ldots i_{n-1}}^{(n-1)}, \tilde{\theta}^{(n-1)}, \eta_n\in[0, \pi]~\forall i_1, \dots i_{n-1}\in\{0, 1\}$. However, before introducing the inequality, we need to check that the rest of the terms have the same sign. To do so, we first sandwich the terms corresponding to the block $n-2$ with the state $\ketzero{1}$ as well:
    \small
    \begin{align}
      &\brazero{n}\mathfrak{U}_n^\dagger \tilde{\mathfrak{U}}_n\ketzero{n} \nonumber\\
      &=\brazero{(n-2)}
      \left(U_0^\dagger(\theta^{(0)})\otimes\iden{(n-3)}\right)\cdots \left(U_{n-3}^\dagger(\bm{\theta}^{(n-3)})\right)\nonumber\\&
      \left(\sum_{i_1, \ldots, i_{n-1} = 0}^1 \ket{i_1\ldots i_{n-2}}\bra{i_1\ldots i_{n-2}}\cdot \bra{0}  R_y^\dagger(\theta_{i_1\ldots i_{n-2}}^{(n-2)})\ket{i_{n-1}}\bra{i_{n-1}}R_y(\tilde{\theta}^{(n-2)})\ket{0} \right.\nonumber\\& \cdot\left.\cos\left(\frac{\theta_{i_1\ldots i_{n-1}}^{(n-1)}-\tilde{\theta}_{i_1\ldots i_{n-1}}^{(n-1)}}{2}\right)\right)
      \cdot \left(U_{n-3}(\bm{\tilde{\theta}}^{(n-3)})\right)\cdots
      \left(U_0(\tilde{\theta}^{(0)})\otimes\iden{(n-3)}\right)
      \ketzero{(n-2)}.
    \end{align}
    \normalsize
    Here, since all the angles are between 0 and $\pi$, we have that
    \small
    \begin{align}
    &\bra{0}R^\dagger_y(\theta_{i_1\ldots i_{n-2}}^{(n-2)})\ket{0}\bra{0}R_y(\tilde{\theta}^{(n-2)})\ket{0}=\cos\frac{\theta_{i_1\ldots i_{n-2}}^{(n-2)}}{2}\cos\frac{\tilde{\theta}^{(n-2)}}{2}\geq 0,\\
      &\bra{0}R^\dagger_y(\theta_{i_1\ldots i_{n-2}}^{(n-2)})\ket{1}\bra{1}R_y(\tilde{\theta}^{(n-2)})\ket{0}=\sin\frac{\theta_{i_1\ldots i_{n-2}}^{(n-2)}}{2}\sin\frac{\tilde{\theta}^{(n-2)}}{2}~\geq 0,
  \end{align}
  \normalsize
  $\forall i_1, \dots i_{n-2}\in\{0, 1\}$. Then, by proceeding analogously with the rest of the blocks, we obtain that all terms are positive. Hence, we can apply the inequality:
  \small
    \begin{align}
      &\brazero{n}\mathfrak{U}_n^\dagger \tilde{\mathfrak{U}}_n\ketzero{n} \nonumber\\
      &\geq \brazero{(n-1)}\left(U_0^\dagger(\theta^{(0)})\otimes\iden{(n-2)}\right)\cdots\left(U_{n-3}^\dagger(\bm{\theta}^{(n-3)})\otimes\mathbbm{1}\right)\nonumber\\
      &\cdot\left(\sum_{i_1, \ldots, i_{n-2} = 0}^1 \ket{i_1\ldots i_{n-2}}\bra{i_1\ldots i_{n-2}}\cdot \otimes \sum_{i_{n-1} = 0}^1 R_y^\dagger(\theta_{i_1\ldots i_{n-2}}^{(n-2)})\ket{i_{n-1}}\bra{i_{n-1}}R_y(\tilde{\theta}^{(n-2)})\right)\nonumber\\
      & \cdot\left(U_{n-3}(\bm{\tilde{\theta}}^{(n-3)}\otimes\mathbbm{1})\right)\cdots \left(U_0(\tilde{\theta}^{(0)})\otimes\iden{(n-2)}\right)\ketzero{(n-1)}\cdot\cos\left({\eta_n/2}\right)\nonumber\\
      &=\brazero{(n-1)}\left(U_0^\dagger(\theta^{(0)})\otimes\iden{(n-2)}\right)\cdots\left(U_{n-3}^\dagger(\bm{\theta}^{(n-3)})\otimes\mathbbm{1}\right)\nonumber\\
      &\cdot\left(\sum_{i_1, \ldots, i_{n-2} = 0}^1 \ket{i_1\ldots i_{n-2}}\bra{i_1\ldots i_{n-2}}\cdot \otimes  R_y^\dagger(\theta_{i_1\ldots i_{n-2}}^{(n-2)})R_y(\tilde{\theta}^{(n-2)})\right)
      \nonumber\\
      &\cdot\left(U_{n-3}(\bm{\tilde{\theta}}^{(n-3)})\otimes\mathbbm{1}\right)\cdots\left(U_0(\tilde{\theta}^{(0)})\otimes\iden{(n-2)}\right)\ketzero{(n-1)}\cdot\cos\left({\eta_n/2}\right)\nonumber\\
      &=\brazero{(n-1)}\mathfrak{U}_{n-1}^\dagger \tilde{\mathfrak{U}}_{n-1}\ketzero{(n-1)}\cdot\cos\left({\eta_n/2}\right).
    \end{align}%}
    \normalsize

      Finally, using the induction hypothesis, we conclude that
      \begin{align}
      \brazero{n}\mathfrak{U}_n^\dagger \tilde{\mathfrak{U}}_n\ketzero{n}
      &\geq \prod_{k=1}^{n-1}\cos\left({\eta_k/2}\right)\cdot\cos\left({\eta_n/2}\right) = \prod_{k=1}^{n}\cos\left({\eta_k/2}\right) \nonumber\\&\implies F \geq \prod_{k=1}^{n}\cos^{2}\left(\eta_k/2\right).
      %\qedsymbol
      \end{align}
  \end{itemize}
\end{proof}

Notice that one of the conditions of the previous Theorem is that $\eta_k\leq\pi$ $\forall k\in\{1, \dots, n\}$. Recall that, as seen in Eq.~(\ref{eq:representative_ap}), $\eta_k\equiv\frac{\delta_k}{8}\eta$. Therefore, we obtain the following condition for the value of $\eta$ for Th.~\ref{th:fid} to be satisfied:
\begin{equation}
  \label{eq:eta}
  \frac{\delta_k}{8}\eta \leq \pi \implies \eta \leq 2^{k-1}\cdot8\pi.
\end{equation}
In particular, if this holds for the smallest $k$, with value $1$, it will be satisfied for the rest of the blocks. Then, we can write the previous condition simply as $\eta\leq 8\pi$.
\subsubsection{Expression of $k_0$}\label{ap:k0}
Assume only clusterization for $k > k_0$:
\begin{align}\label{eq:1}
    F \geq \prod_{k=k_0+1}^n \cos^2\left(\frac{\eta}{4\cdot2^k}\right) \geq \prod_{k=k_0+1}^n e^{-2\left(\frac{\eta}{4\cdot2^k}\right)^2} = e^{-2\frac{\eta^2}{16}\sum_{k=k_0+1}^n4^{-k}},
\end{align}
since $\cos(x) \geq e^{-x^2}$ for $x\lessapprox\pi/2$. Also, the sum is a geometric series with value
\begin{equation}
    \sum_{k=k_0+1}^n4^{-k} = \frac{1}{3}\left(4^{-k_0}-4^{-n}\right).
\end{equation}
Then,
\begin{align}
    F \geq e^{-\frac{2}{3}\frac{\eta^2}{16}\left(4^{-k_0}-4^{-n}\right)} \coloneqq F_{k_0}.
\end{align}
Let us finally find an expression for $k_0$ given minimum fidelity $F_{k_0}=1-\epsilon$:
\begin{align}
    -\frac{2}{3}\frac{\eta^2}{16}\left(4^{-k_0}-4^{-n}\right) = \log{F_{k_0}}\\
    \left(4^{-k_0}-4^{-n}\right) = -\frac{3}{2}\frac{16}{\eta^2}\log{F_{k_0}}\\
    4^{-k_0} = 4^{-n}-\frac{3}{2}\frac{16}{\eta^2}\log{F_{k_0}}\\
    -k_0\log{4} = \log{\left(4^{-n}-\frac{3}{2}\frac{16}{\eta^2}\log{F_{k_0}}\right)}\\
    k_0 = -\frac{\log{\left(4^{-n}-\frac{3}{2}\frac{16}{\eta^2}\log{1-\epsilon}\right)}}{\log{4}}.\label{eq:k0s}
\end{align}

Note that in the limit for both $F_0\to 1$ and $\eta\to\infty$, we have that
\begin{equation}
  k_0\to -\log(4^{-n})/\log(4)=n,
\end{equation}
as expected.

Also, since $k_0$ must be an integer, we take the ceiling of its value. In addition, we consider the minimum of this parameter to be 2, so the condition of $\eta\leq 8\pi$ is satisfied. Hence, the final expression for $k_0$ is

\begin{equation}
  \label{eq:k0_ap2}
k_0 = \max \left \{ \lceil -\frac{1}{2}\log_2 (4^{-n}-\frac{96}{\eta^2}\log (1-\epsilon))\rceil,2 \right \},
\end{equation}
where we have defined $\epsilon \equiv 1-F_0$.

Once we have a fixed number of uniformly controlled rotation blocks, $k_0$, for which there is no clustering, we count the number of necessary gates. The gate $F_{k}^{k-1}(\bm{\hat{y}}, \bm{\theta})$ can be implemented with $2^{k-1}$ CNOTs and $2^{k-1}$ single-qubit gates~\cite{MIKKO3GATES}. Thus, the total number of gates is:
\begin{align}
  &\# \text{CNOTs} = \sum_{k=1}^{k_0}2^{k-1}=2^{k_0}-1,\label{eq:cnot-k0}\\
  &\# \text{SQGs} \;\;\;= \sum_{k=1}^{k_0}2^{k-1} + \sum_{k=k_0+1}^{n}1 = 2^{k_0}-1 + n-k_0.
\end{align}
%%%%%%%%%%%%%%%%%%%%%%%%%%%%%%%%%%%%%%%%%%%%%

\begin{figure}[t!]
  \centering
  \includegraphics[width=0.86\columnwidth]{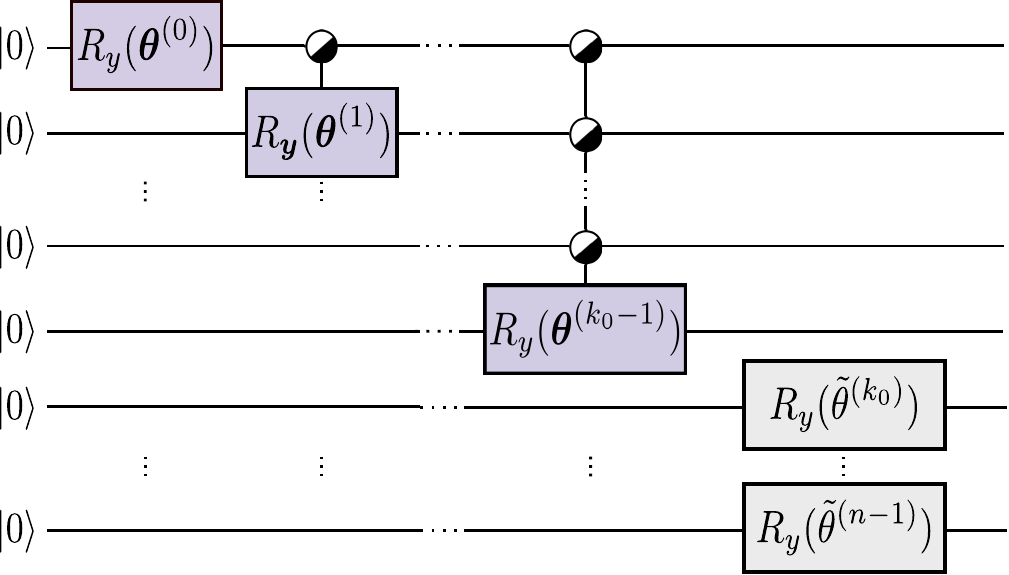}
  \caption{Quantum circuit performing the protocol presented in this section, based on the Grover-Rudolph method for a system of $n$ qubits.}
  \label{Fig:Circuit}
\end{figure}

\subsection*{B.3 Final Protocol}
%\subfile{sections/appendix/protocol}
\begin{de1}\label{df:f}
Let $f\colon [x_\text{min}, x_\text{max}]\to ${\rm I\!R}$^{+}$ be a positive function in $L^2([x_\text{min}, x_\text{max}])$. We define the $n$-qubit normalized representative state of $f(x)$ as the $n$-qubit state $|f(x)\rangle_n=\sum_{j=0}^{2^n-1} f(x_\text{min}+j\delta_n)|j\rangle$, with $\delta_n=\frac{x_\text{max}-x_\text{min}}{2^n-1}$ and $\sum_{j=0}^{2^n-1}f^2(x_\text{min}+j\delta_n) = 1$.
\end{de1}

\begin{th1}
\label{th:k0}
Let $f:[0,1]\rightarrow${\rm I\!R}$^+$ be a positive integrable function in $L^2([0,1])$ and $0\leq\eta\leq 8\pi$ a constant such that $\eta=\sup_{x\in[0,1]}\left|\partial^2_x \log f^2(x) \right|$. Then, it is posible to approximate the $n$-qubit representative state of $f(x)$, $|f(x)\rangle_n$, by a quantum state $|\Psi(f)\rangle_n$ such that the fidelity $|\langle \Psi(f)|f(x)\rangle_n|^2 \geq 1-\epsilon$ with at most $2^{k_0(\epsilon)}-1$ two-qubit gates, with
\begin{equation}\label{eq:k0_ap}
k_0(\epsilon) = \max \left \{ \lceil -\frac{1}{2}\log_2 (4^{-n}-\frac{96}{\eta^2}\log (1-\epsilon))\rceil,2 \right \},
\end{equation}
and the circuit to perform it is given in Fig.~\ref{Fig:Circuit}.

\end{th1}
\begin{proof}
Using Th.~\ref{th:2derlog} and Cor.~\ref{cor:rep}, we can define the representative angle for each block as in Eq.~(\ref{eq:representative_ap}). Then, by applying Th.~\ref{th:fid} with the development between Eqs. (\ref{eq:1}) and (\ref{eq:k0s}), it is clear that the fidelity will be larger than or equal to $1-\epsilon$. Finally, Eq.~(\ref{eq:cnot-k0}) states that the number of two-qubit gates necessary to realize this protocol is $2^{k_0}-1$, as we wanted to prove.
\end{proof}
%%%%%%%%%%%%%%%%%%%%%%%%%%%%%%%%%%%%%%%%%%%%%

%%%%%%%%%%%%%%%%%%%%%%%%%%%%%%%%%%%%%%%%%%%%%
%\twocolumngrid

\end{document}